\documentclass[12pt]{article}
\usepackage[dvips]{epsfig}
\usepackage{url}
\usepackage{comment}

\makeatother
\usepackage[sort]{natbib}
\bibpunct{(}{)}{;}{a}{}{,}

 \usepackage{graphicx,fullpage,amsmath,multicol,multirow,amssymb,amsbsy,pifont,subfig}
 \usepackage{setspace}
 \usepackage{epsfig}
 \usepackage{amsmath, amsthm, amssymb, bm}
\numberwithin{figure}{section}
\numberwithin{table}{section}
\numberwithin{equation}{section}
 \usepackage{color}
\usepackage{graphics}
\usepackage{booktabs}
 \usepackage{wrapfig}

\newcommand{\rowgroup}[1]{\hspace{-1em}#1}
\newcommand{\bfX}{\boldsymbol{X}}
\newcommand{\bfR}{\boldsymbol{R}}
\newcommand{\bfY}{\boldsymbol{Y}}

\newcommand{\bfD}{\boldsymbol{D}}
\newcommand{\bfeta}{\boldsymbol{\eta}}
\newcommand{\bfbeta}{\boldsymbol{\beta}}
\newcommand{\bfalpha}{\boldsymbol{\alpha}}
\newcommand{\bfepsilon}{\boldsymbol{\epsilon}}
\newcommand\independent{\protect\mathpalette{\protect\independenT}{\perp}}
\def\independenT#1#2{\mathrel{\rlap{$#1#2$}\mkern2mu{#1#2}}}

\title{Bayesian Profiling Multiple Imputation for Missing Electronic Health Records}
\author{Yajuan Si$^{1,*}$, 
Mari Palta$^{2}$, and Maureen Smith$^{2}$\\
$^{1}$University of Michigan, Ann Arbor, Michigan, U.S.A.\\
$^{2}$University of Wisconsin-Madison, Madison, Wisconsin, U.S.A.\\
$^*$email: yajuan@umich.edu}

\begin{document}
\maketitle

\begin{abstract}
Electronic health records (EHRs) are increasingly used for clinical and comparative effectiveness research, but suffer from missing data. Motivated by health services research on diabetes care, we seek to increase the quality of EHRs by focusing on missing values of longitudinal glycosylated hemoglobin (A1c), a key risk factor for diabetes complications and adverse events. Under the framework of multiple imputation (MI), we propose an individualized Bayesian latent profiling approach to capture A1c measurement trajectories subject to missingness. The proposed method is applied to EHRs of adult patients with diabetes in a large academic Midwestern health system between 2003 and 2013 and had Medicare A and B coverage. We combine MI inferences to evaluate the association of A1c levels with the incidence of acute adverse health events and examine patient heterogeneity across identified patient profiles. We investigate different missingness mechanisms and perform imputation diagnostics. Our approach is computationally efficient and fits flexible models that provide useful clinical insights.  

\end{abstract}

\section{Introduction}

\subsection{Glycemic testing and control}

Diabetes is a condition requiring intensive management, and a major cause of morbidity and mortality. Approximately 25.2\% of American seniors have diabetes and would benefit from individualized guidelines on glycemic testing and control~\citep{CDC-diabetes17}, but little clinical evidence exists to develop such plans. Current evidence-based guidelines encouraging tight glycemic control are most applicable to relatively healthy patients with diabetes. Glycemic control is usually monitored through glycosylated hemoglobin (A1c), which reflects blood glucose values over approximately the last 3 months. The American Diabetes Association (ADA) recommends A1c testing at least two times a year in patients who are meeting treatment goals and have stable glycemic control, and quarterly in patients whose therapy has changed or who are not meeting glycemic goals~\citep{diabetes18-target}. Individuals without diabetes have A1c values below 5.7\%, and maintaining A1c below 7\% is recommended for individuals with diabetes to avoid chronic diabetes complications~\citep{diabetes18-target}.

For patients aged 65 years or over and those with comorbid conditions, adhering tightly to guidelines established for those younger and healthier, might be difficult and diminish the quality of life. Tight control can lead to acute adverse outcomes, such as hypoglycemic coma, seizures, falls, fractures, motor vehicle accidents, cardiovascular events, stroke, and acute renal failure~\citep{negativeoutcome-shorr97,negativeoutcome-allen01,negativeoutcome0kennedy02,negativeoutcome-davis04,negativeoutcome-boyd05,negativeoutcomes-pogach07,negativeoutcome-qaseern07,negativeoutcome-schech07}. For example, the ACCORD and ADVANCE trials find that lowering A1c values increases the risk of cardiovascular death~\citep{longtermoldwithCVD-martin06,va-trials-dluhy08}. The relationship between A1c and adverse outcomes in such patients may be U-shaped, and less stringent A1c goals may be appropriate~\citep{diabetes18-older}.

To supplement findings from clinical trials, electronic health records (EHRs) are increasingly a data source for clinical and comparative effectiveness research on improving health care~\citep{ehr:nejm:cebul11,ehr:ahrq13}. The EHR provides a longitudinal record of patient medical information that is maintained by encounters in any care delivery setting, and includes key clinical data relevant to patient care, such as laboratory data (e.g., A1c values), demographics, progress notes, problems, medications, vital signs, past medical history, immunizations and radiology reports. Linked insurance data, for example from Medicare~\citep{cms:ehr}, provide information on adverse clinical events through billed claims. However, the quality of EHRs is reduced by a large amount of intermittent missing data, because data are typically collected in an unscheduled fashion when the patient seeks care or the physician orders care.

In this article, we aim to use EHR to further understand the relationship between A1c levels and acute adverse outcomes. Information on adverse outcomes is rarely missing, as it is obtained from billing records. However, A1c within 3 months prior to an event may not be available due to the above recommendations. Multiple additional factors could influence the number of missing A1c values for individual patients. Physician non-adherence to guidelines for lab testing will cause missing values. The patient's health potentially affects the propensity to test, leading to informative missingness patterns that may cause systematic estimation bias. As argued by~\cite{Sebastien:Daniels16}, the complex interplay between health care providers and patients could result in various missing data mechanisms. Incorporation of all available information including patient health status is critical to appropriately handle missing A1c values from EHRs. Furthermore, capturing A1c trajectories and their relationship to patient characteristics is of clinical interest for health risk prediction and guideline establishment.

We propose a Bayesian profiling approach to incorporate patient characteristics and infer latent groups of A1c measurement trajectories. The measurement trajectories reflect both the A1c collection patterns and A1c values across time. The Bayesian profiling approach is combined with multiple imputation (MI,~\cite{rubin:1987}) to produce complete datasets for general analysis purposes. As an illustration of the MI inference, we evaluate the association between A1c levels and the incidence of any acute health events, such as hospitalization, emergency room (ER) visits or death. The main novelty lies in generalizing MI with a flexible imputation engine, a covariate-dependent latent profile model, to depict nonlinear longitudinal trajectories and incorporate numerous covariates into the latent profiling. 

\subsection{Statistical literature on intermittently missing data}

Missing A1c values in EHRs present statistical methodology challenges. Intermittent missing data in large-scale, unbalanced longitudinal studies, where subjects reappear after one or more missed visits, call for new imputation approaches especially when the missing values are potentially nonignorable~\citep{littlerubin} and the observed cases are sparse. Simple methods (e.g., complete case analysis or last observation carried forward) will distort the relationship and are not recommended by the~\cite{nrc10}. Standard missing data methods in longitudinal studies focus on a common set of pre-specified and monotone missing times, where a measurement being missing implies that all follow-up measurements are also missing or dropped from the analysis (e.g., reviews in~\cite{ibrahim:review09}). Likelihood-based, weighting or factorization approaches for missing data mainly apply to monotone missingness~\citep{daniels2008}. Utilization of the information collected after subjects reappear will potentially correct bias and increase estimation efficiency, and the challenge lies in how to appropriately model sparse data structures with non-monotone missing patterns. Weighting adjustment for longitudinal data with intermittent missingness is complex and computationally demanding, and there is no consistent recommendation on how the weights should be included in the longitudinal data modeling~\citep{littlerubin}. \cite{Tchetgen:JASA16} consider inverse probability weighting for non-monotone missing at random (MAR) data, whereas we argue that MI provides a superior solution to coherently utilize all available information and offer flexibility for model building with broad analytic goals. 

We seek a flexible MI engine to predict intermittent missing values. MI separates imputation and analysis into two steps and propagates the uncertainty due to missing data. Various MI software packages have been developed assuming data are MAR and based on either joint multivariate normal distributions, e.g., PROC MI~\citep{procmi15}, Amelia~\citep{amelia} and norm~\citep{schafer:1997}, or a sequence of fully conditional distributions, e.g., IVEware~\citep{raghu:2001}, mice~\citep{oos}, and mi~\citep{mi-manual15}. Multilevel models are embedded with MI to handle correlated data, such as the packages REALCOM-IMPUTE~\citep{carpenter-missbook} and pan~\citep{pan16}. However, existing MI software cannot handle high-dimensional data that are subject to high proportions of intermittent missingness in large-scale longitudinal studies, nor nonignorable missing values~\citep{si:reiter:12}.

MI for missing not at random (MNAR) data requires a joint model for the incomplete variables (i.e., the A1c values) and the missingness indicators (i.e., whether A1c values are present on any given occasion). Examples of joint models include selection models, pattern-mixture models, and shared parameter models~\citep{caroll88,little1995}. Assumptions have to be introduced for parameter identification, which cannot be verified with the available observations, and the computation is non-trivial~\citep{ibrahim:chen:lip:herr}. The longitudinal A1c collection history in EHRs results in sparse observations across a large number of different missingness patterns and calls for flexible modeling strategies that account for the time dependency, borrow information across patterns and capture patient heterogeneity. 

We jointly model the A1c missingness patterns and the lab values taking into account patient characteristics. To achieve parameter identification, dimension reduction and straightforward interpretation, we introduce latent profiles and develop a Bayesian profiling multiple imputation (BPMI) approach. The novel latent profiling can handle ignorable and nonignorable missingness under the conditional independence assumption given the latent profiles, which yields a consistent estimation of parameters that are not associated with the latent profiles~\citep{McCulloch16}.

We assume that latent profiles are primarily determined by the trajectories of the longitudinal A1c measurements, including A1c values and potentially the measurement patterns. The latent profiling includes patient characteristics---such as socio-demographics, healthcare utilization measures, medications, medical complexity indicators, comorbidity and complications---as covariates that affect the profile assignment. This is an improvement over previous work that required joint modeling of the outcome variables, missingness indicators and covariates, leading to the need for Monte Carlo integration or approximate inferences (e.g., \cite{roy-edp18}). The proposal shares a similar decomposition with~\cite{lin2004} but differs by the MI framework under the Bayesian paradigm. Besides avoiding the extra computation step of Monte Carlo integration, the posterior updating of BPMI has efficiency gains via Gibbs sampling and generates completed datasets for general analysis purposes. 

As important contributions, the paper uses all available information with intermittent missingness to infer latent groups that are relevant to substantive interpretations, fits flexible mixture models that can capture irregular data distributions and develops Gibbs samplers to achieve computational efficiency. We examine the identified profiles and present the profile-specific characteristics to describe the subgroup heterogeneity. 

The paper is organized as follows. Section~\ref{data} describes the EHR structure and content. The BPMI method is presented in detail in Section~\ref{methods}. We compare the new method with alternative imputation methods via simulation studies in Section~\ref{simulation} and apply it to the EHRs in Section~\ref{application}, where our statistical and substantive findings are presented. Section~\ref{discussion} presents contributions and discusses future extensions.

\section{EHR description} 
\label{data}

\noindent We extract the EHR data of adult patients with diabetes who belonged to a large academic Midwestern health system between 2003 and 2013. Patients are included if they have diabetes defined by a qualifying International Classification of Diseases-9 code for diabetes. We treat the first four quarters after the initial enrollment in the health care system as the baseline period (time 0). For inclusion, patients must have baseline A1c values and at least one available follow-up A1c measurement. Patients are excluded from the analysis if they do not have continuous Medicare Part A \& B fee-for-service, or if they are not seen at a clinic located in the system with access to laboratory data. 

Because A1c reflects mean blood glucose over the preceding three months and quarterly testing is recommended for many patients, we construct a longitudinal dataset with one measurement per patient per three-month period (patient-quarter). When there are multiple A1c values available during the three months, we use the average of the measurements before the date of first acute health event (e.g., hospitalization, ER visit or death) date if there are any such events. 

Missingness occurs when no measurements are available, or as occurs in a handful of cases (0.3\%) after the first acute event in the quarter.  We use the average of collected A1c values during the first four quarters as the baseline A1c value. Patient eligibility for inclusion is evaluated each quarter and patients are followed until loss to follow-up (e.g., no longer in the health system) or death, where for patients who die during the study period, we keep the patient in the dataset until after the quarter of death. Patients' various enrollment dates and lengths of follow-up result in an unbalanced data structure. We identified 7372 patients with 113761 quarters after baseline and only 57285 available observations. The proportions of intermittent missingness across patients are as high as 0.97, with a median of 0.50. The baseline A1c values of 7372 patients center around mean 6.93\% and range from 3.55\% to 17.1\%, showing heterogeneity in glycemic control.  

\begin{figure}
\begin{tabular}{c}
\includegraphics[width=0.95\textwidth]{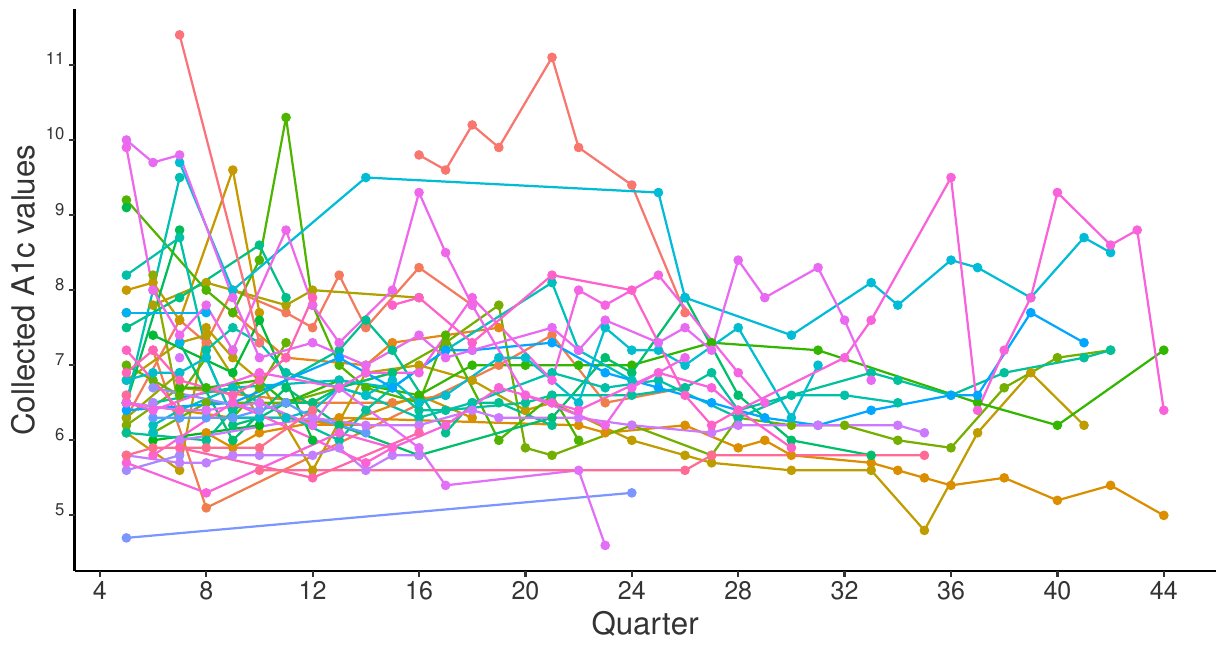}
\end{tabular}
\caption{Observed A1c values of 30 randomly selected patients. Each line represents one patient, with segments connecting two available subsequent A1c measurements.}
\label{a1c_30}
\end{figure}

A ``spaghetti" plot of A1c values is presented for 30 randomly selected patients in ~\ref{a1c_30}. Time trends and trajectories vary between patients, and repeated measurements of the same patients tend to be correlated. We consider mixed-effects models accounting for within-patient correlation with random intercepts and slopes with respect to time or functions of time. While the A1c measurement trajectories present patient heterogeneity, subgroups of patients could share similar profiles across time. We assume each subgroup to have different location and scale parameters for the time trends, which results in a mixture distribution overall.

Our goal is to estimate the A1c trajectories over time and identify latent A1c pattern groups, where the patient-level covariates affect the group allocation probabilities. The latent profiling allocation in the model takes into account variables that predict the A1c values or missingness patterns. EHRs collect a rich set of patient characteristics that could predict the A1c trajectories and likelihood of missingness patterns. The patient-quarter structure creates both time-invariant and time-varying variables. 

Baseline characteristics include socio-demographics, healthcare utilization measures, medications, complexity, comorbid conditions~\citep{elixhauser-comorbidity98}, and diabetes complications, all of which are carried forward (e.g., chronic conditions will be assumed as present from diagnosis onward). The comorbid patterns are characterized by Hierarchical Condition Categories (HCC) risk scores. The number of adverse events, such as hospitalization and ER visits, and the number of ER visits that did not lead to hospitalization during the first four quarters, serve as proxy measures of healthcare utilization. A large number of variables and low prevalence of incidence cause problems in the covariate-dependent allocation, a well-known problem for latent class analysis~\citep{Vermunt10}. We thus create a summary variable for each patient capturing the count of conditions with prevalence $<2\%$. The covariates themselves could have high co-linear dependence and structural constraints. We use the observed A1c values ignoring the patient-quarter structure and fit an ordinary linear regression to select the variables that are highly correlated with A1c. We then fit a logistic regression model to predict the missingness at each quarter and select the variables that are predictive of the missingness pattern.

\begin{table}
\centering
\footnotesize
\caption{Summary of baseline characteristics: mean (standard deviation) and percentage.}
\label{predictor}
\begin{tabular}{ll}
\\\toprule  
\rowgroup{\em Sociodemographics}&\\
Age & 69.91 (10.76) \\ 
Female & 53\% \\ 
White&93\%\\
Medicaid & 14\% \\ 
  \rowgroup{\em Utilization}&\\
 Adverse event count & 0.58 (0.88) \\ 
Hierarchical condition categories & 1.41 (0.98) \\  
 % bl\_gfr\_imp & 68.83 (21.27) \\ 
 %  bl\_ed\_visits\_cnt 
 % that did not lead to hospitalization
Number of  emergency room visits & 0.12 (0.37) \\ 
  % v\_ac\_clm\_v\_cnt & 2.10 (2.06) \\ 
  \rowgroup{\em Comorbidity and complication}&\\ 
%  c\_elix\_low\_prev\_cnt 
Count of low prevalence conditions&0.21 (0.54)\\
%c\_amp\_newton\_bf\_coh\_ind & 0.01 \\ 
 % c\_ckd\_foley\_bf\_coh\_ind 
 Chronic kidney disease & 14\% \\ 
  %c\_cpd\_elix\_bf\_coh\_ind 
  Chronic pulmonary disease& 18\% \\ 
  %c\_dem\_taylor\_bf\_coh\_ind 
  Dementia& 5\% \\ 
  %c\_dep\_elix\_bf\_coh\_ind 
 Depression & 16\% \\ 
  %  c\_ent\_disability\_bf\_coh\_ind 
  Entitlement disability & 18\% \\ 
% c\_eye\_newton\_bf\_coh\_ind 
  Eye disease& 14\% \\ 
 % c\_htn\_tu\_bf\_coh\_ind 
  Hypertension& 85\% \\ 
  %c\_hypothy\_elix\_bf\_coh\_ind 
 Hypothyroidism& 14\% \\ 
%  c\_lytes\_elix\_bf\_coh\_ind 
Fluid and electrolyte disorders & 14\% \\ 
      \end{tabular}
  \begin{tabular}{ll}
%  c\_neuro\_elix\_bf\_coh\_ind 
  Other neurological disorders& 7\% \\ 
 % c\_obese\_elix\_bf\_coh\_ind 
 Obesity& 16\% \\ 
  %c\_pcirc\_elix\_bf\_coh\_ind & 0.03 \\ 
 % c\_psych\_elix\_bf\_coh\_ind 
  Psychoses& 9\% \\ 
  %c\_pvd\_elix\_bf\_coh\_ind 
%  Peripheral Vascular Disease& 12\% \\ 
 % c\_pvd\_newton\_bf\_coh\_ind
 % Peripheral Vascular Disease & 23\% \\ 
    %c\_renal\_elix\_bf\_coh\_ind 
   Renal failure & 7\% \\ 
%  c\_tumor\_elix\_bf\_coh\_ind 
  Solid tumor without metastasis& 6\% \\ 
  %c\_ulcle\_newton\_bf\_coh\_ind 
  Lower extremity ulcer& 7\% \\ 
  %c\_valv\_elix\_bf\_coh\_ind 
  Valvular disease & 8\% \\
%  bl\_gfr\_avg: $\geq$ 90&0.16\\
%  bl\_gfr\_avg: 60 - 90&0.51\\
%  bl\_gfr\_avg: 30 - 60&  0.31\\
%  bl\_gfr\_avg: 15 - 30&  0.02\\
%  bl\_gfr\_avg: $\leq$15& 0.00\\
  Kidney damage-light & 16\%\\ 
  Kidney damage-mild & 51\%\\ 
  Kidney damage-moderate & 30\% \\ 
  Kidney damage-severe & 3\% \\ 

% \rowgroup{\em Complexity}&\\
Congestive heart failure&19\%\\
Ischemic heart disease&21\%\\
%None&0.60\\
\rowgroup{\em Medication prescribed}&\\
Insulin&13\%\\
Sulfonylureas&17\%\\
Hypoglycemics&16\%\\
Other&54\%\\
\bottomrule
\end{tabular}
\end{table}

The union of the two sets of selected variables results in the covariate list in Table~\ref{predictor}. The baseline characteristics are fully observed without missing values. The table shows that the cohort has a modest number of complex patients. For example, 14\% of patients have Chronic Kidney Disease, 33\% suffer from moderate or severe kidney damage, 19\% have Congestive Heart Failure, 21\% have Ischemic Heart Disease, and 85\% have hypertension. 

Time-varying variables include age and the count of physician visits that occurred within each quarter. On average patients have 2.39 physician visits every three months. We exclude time-varying Body Mass Index (BMI), blood pressure (BP), and elevated low-density lipoprotein (LDL), previously shown as associated with tight A1c control~\citep{bmi-risk-mcfarlan02,emvisits-niefeld03,emvisits-jackson06} but subject to high missingness. LDL is often recommended for annual measurements. Measures such as BP and BMI are usually available when the patient sees a provider in a face-to-face visit, which occurs at A1c tests. The majority of BP records before 2007 in our dataset tend to be missing. Hence, the missingness percentages of these time-varying covariates are all above 50\%, similar to those of A1c values. 

Missing data in the time-varying covariates cause computational problems, and here we only consider missing A1c values and include other variables that are fully collected. The extension to handle missing values in the baseline and time-varying variables and perform a systematic variable selection by propagating all sources of uncertainty is discussed in Section~\ref{discussion}.

\section{Bayesian profiling multiple imputation}
\label{methods}

Denote the individual record for patient $i$ by $\{\bfX_{i0}, \bfX_{ij}, Y_{ij}, R_{ij} \mbox {, }j\in[1,T_i]\}$ for the total number $T_i$ of tracked quarters during the follow-up, where $\bfX_{i0}$ are the time-invariant covariates, $\bfX_{ij}$'s are time-varying covariates, and $Y_{ij}$ is the variable to be imputed, A1c values over time. For brevity, we will refer to $Y_{ij}$ as the longitudinal outcome. Assume that only $Y_{ij}$'s are subject to missing values, and let $R_{ij}$ be a time-varying binary indicator for its response, $R_{ij}=1$ if $Y_{ij}$ is observed, $R_{ij}=0$ otherwise. Denote the time-varying variables by $\bfY_i=(Y_{i1}, \dots,Y_{iT_i})^\intercal$, $\bfX_i=(\bfX_{i1}, \dots,\bfX_{iT_i})^\intercal$, and $\bfR_i=(R_{i1}, \dots,R_{iT_i})^\intercal$, for $i=1,\dots, n$ total number of patients. 

Assume individuals fall into latent classes $C_i=1,\dots,L$, where $L$ is a finite and positive known integer as the total number of classes. The selection of $L$ and the case of $L$ being an unknown random variable that induces additional uncertainty are discussed in Section~\ref{application} and Section~\ref{discussion}. We assume that the latent class structure is primarily determined by the trajectories of the longitudinal outcome and potentially its missingness pattern, and that the allocation probabilities are affected by the covariates. 

The conditional distribution of the measurements given observed data can be expressed as
\begin{align*}
f(\bfY_i|\bfX_{i0}, \bfX_{i}, \bfR_i=1)%&=\sum_{l}f(\bfY_i|\bfX_i,\bfR_i=1,C_i=l)f(C_i=l|\bfX_{i},\bfR_i=1)\\
&=\sum_{l=1}^L\frac{f(\bfR_i=1|\bfY_i,\bfX_{i0}, \bfX_i,C_i=l)f(\bfY_i|\bfX_{i0}, \bfX_i,C_i=l)f(C_i=l|\bfX_{i0}, \bfX_i)}{f(\bfR_i=1|\bfX_{i0}, \bfX_i)},
\end{align*}
where $f(\cdot)$ denotes the distribution.
%\begin{align*}
%f(Y_{ij}|\bfX_{i}, R_{ij}=1)&=\sum_{l}f(Y_{ij}|\bfX_i,R_{ij}=1,C_i=l)f(C_i=l|\bfX_{i},R_{ij}=1)\\
%&=\sum_{l}\frac{f(R_{ij}=1|Y_{ij},\bfX_i,C_i=l)f(Y_{ij}|\bfX_i,C_i=l)f(C_i=l| \bfX_i)}{f(R_{ij}=1| \bfX_i)},
%\end{align*}

If $\bfR_i$ only depends on fully observed $(\bfX_{i0},\bfX_{i})$, i.e., missingness is MAR, the observed data can provide valid inference since $f(\bfY_i|\bfX_{i0},\bfX_{i}, \bfR_i=1)=f(\bfY_i|\bfX_{i0},\bfX_{i})$. The profile structure is then independent of the missingness patterns.
\begin{align*}
f(\bfY_i|\bfX_{i0}, \bfX_{i}, \bfR_i=1)%&=\sum_{l}f(\bfY_i|\bfX_i,\bfR_i=1,C_i=l)f(C_i=l|\bfX_{i},\bfR_i=1)\\
&=\sum_{l=1}^Lf(\bfY_i|\bfX_{i0}, \bfX_i,C_i=l)f(C_i=l|\bfX_{i0}, \bfX_i).
\end{align*}

We call this marginal profiling because the latent class structure influences only the marginal distribution of $Y$. The MAR assumption can be relaxed to
\begin{align}
\label{joint}
f(\bfR_i=1|\bfY_i,\bfX_{i0},\bfX_i,C_i=l)=f(\bfR_i=1|\bfX_{i0},\bfX_i,C_i=l),
\end{align}
that is, conditional MAR, where the missingness is independent of the outcome given the latent classes and covariates. However, the unconditional missingness is nonignorable as the latent structure affects both the longitudinal outcome and missingness patterns. Joint modeling of $(\bfY_i,\bfR_i)$ is then necessary and will be identified due to the conditional independence assumption given the latent classes. We call this approach joint profiling.

Joint profiling is identical to marginal profiling when the parameters in Model~\eqref{joint} do not change across profiles, so $R_i$ is conditionally independent of $C_i$ given $(\bfX_{i0},\bfX_i)$. Hence, the joint profiling approach can model both ignorable and nonignorable missing data.
We will consider both the marginal and joint profiling for imputation and make inferences on the effect of A1c levels on acute health outcomes. 

The latent classes are at the individual level and thus time invariant. Further, we assume the latent class allocation depends only on time-invariant variables $f(C_i=l| \bfX_{i0},\bfX_i)=f(C_i=l|\bfX_{i0})$. For computational and interpretational convenience, we assume that the outcome trajectory is independent of time-invariant covariates given the latent class $f(\bfY_i|\bfX_i,\bfX_{i0},C_i=l)=f(\bfY_i|\bfX_{i}, C_i=l)$. That implies that the patient characteristics at baseline are represented by the latent profiling.

Here, the conditional independence assumption between $Y_{ij}$ and $R_{ij}$ given $C_i$ is used for identification, and the conditional independence assumptions involving $\bfX_{i0}$ and $\bfX_{ij}$ are introduced for interpretation and can be tested in the model fitting. 

Are our assumptions regarding the missing data too strong? \cite{Sebastien:Daniels16} list various sub-mechanisms---potentially MAR or MNAR---that are relevant to missing EHRs, such as enrollment status, encounters with the health system, measurements and structural changes. \cite{McCulloch16} evaluate biased and unbiased estimation in longitudinal studies with informative missingness and recommend mixed-effects models. In addition to random effects, we introduce latent profiles with a large number of covariates, and expect the flexible latent profiling with rich characteristics to capture the underlying heterogeneity and make the assumptions at least plausible. We rigorously assess the imputation performances in Section~\ref{application}.

\subsection{Marginal profiling}

Under MAR we construct latent profile models based on the collected outcomes, i.e., marginal profiling, which is estimated from the pattern of trajectories of the longitudinal outcomes.

\begin{align}
\label{mar}
f(\bfY^{obs}_i|\bfX_{i0},\bfX_{i})=\sum_{l=1}^Lf(\bfY^{obs}_i|\bfX_{i},C_i=l)f(C_i=l|\bfX_{i0}).
\end{align}
Here $\bfY^{obs}_i$ denotes the observed A1c measurement for individual $i$. The factorization involves an outcome model and a latent profile model. The model specification in~\eqref{mar} assumes: 1) the missingness pattern is independent of the A1c values and the latent profiles given the covariates $R_{ij}\independent (Y_{ij}, C_i)|(\bfX_{i0},\bfX_{ij})$; 2) the A1c values are independent of the missingness patterns and the baseline characteristics given the latent profiles and time-varying covariates $Y_{ij}\independent (R_{ij},\bfX_{i0})|(C_i,\bfX_{ij})$; and 3) the latent profiles are independent of the time-varying covariates given the baseline characteristics $C_{i}\independent \bfX_{ij}|\bfX_{i0}$. Since the role of $\bfX_{i0}$ can be checked, Model~\eqref{mar} can be generalized by assuming the outcome depends on the baseline characteristics and that these are captured by the latent profiling in the current specification. Our sensitivity analysis in the EHR application does not find evidence against the specification with $\bfX_{i0}$, as the covariates affect the allocation assignment probability of different profiles.

In the latent profile model, denote by $f(C_i=l|\bfX_{i0})\doteq \pi_l(\bfX_{i0})$ the allocation probability of pattern $l$ conditional on $\bfX_{i0}$, for $l=1,\dots,L$. We consider a multinomial logistic regression model with coefficient vector $\bfeta_l$, for $l=1,\dots, L$. Set $\bfeta=(\bfeta_1,\dots, \bfeta_L)$ and $\bfeta_1=\vec{0}$ for identification, we have
\begin{align}
\label{pi}
\pi_l(\bfX_{i0})=\frac{\exp(\bfX_{i0}^T\bfeta_l)}{\sum_{k=1}^L\exp(\bfX_{i0}^T\bfeta_k)}.
\end{align}

For posterior computation we develop a Gibbs sampler by introducing P\'{o}lya-Gamma (PG) distributed variables $w_{il}$~\citep{logit-pg13}, $$w_{il}|(\bfX_{i0},\bfeta) \sim \textrm{PG}(1,r_{il}),$$ where $r_{il}=\bfX_{i0}^T\bfeta_l-\log[\sum_{k\neq l}\exp(\bfX_{i0}^T\bfeta_k)]$. Conditional on the PG variables, the posterior distribution of $\bfeta_l$ will be conjugate with normal prior distributions. The resulting Gibbs sampler improves the posterior fitting and outperforms rejection sampling methods with quick convergence.

Let $c_{il}=I(C_i=l)$ be the latent class indicator and introduce the normal prior distribution $\textrm{N}(\mathbf{b}_l,\mathbf{B}_l)$ on the coefficients $\bfeta_l$, the conditional posterior distribution of $\bfeta_l$ given $w$ is multivariate normal, $\pi(\bfeta_l\mid-)\sim \textrm{N}(\mathbf{m}^*_{l},\mathbf{S}^*_{l})$. Here
$\mathbf{S}^*_{l}=(\mathbf{V}^T\Omega_{l}\mathbf{V}+\mathbf{B}_l^{-1})^{-1}$, $\mathbf{V}$ is the design matrix with each row $\bfX_{i0}^T$, $\Omega_{l}=diag(w_{1l},\dots,w_{nl})$, and $\mathbf{m}^*_l=\mathbf{S}^*_l(\mathbf{B}_l^{-1}\mathbf{b}_l+\mathbf{S}^T\mathbf{m}_l)$, where $\mathbf{m}_l$ is a vector in $R^n$, and the $i$th component is $m_l^{(i)}=c_{il}-1/2 + w_{il}\{\log[\sum_{k\neq l}\exp(\bfX_{i0}^T\bfeta_k)]\}$.

For the collected longitudinal and continuous A1c measurements, we assume a linear mixed-effects model with profile-specific coefficients and variances ($\bfbeta^*_C, \sigma^2_C$). Let $\bfD_i$ denote a $T_i\times p$ covariate matrix, with associated $p$-vector of coefficients $\bfbeta$. The $j$th row of $\bfD_i$, denoted by $\bfD_{ij}$, is then a $p$-vector of covariate values measured at time $j$. Covariates for profile-specific effects $\bfbeta^*_{C_i}$ and for individual-specific random effects $\mathbf{b}_i$ are denoted by the $T_i\times q$ matrix $\bfD_i^*$ and the $T_i\times r$ matrix $\bfD_i^{**}$, respectively, both with structures similar to $\bfD_i$. There may be overlap of the covariates in $\bfD_i$, $\bfD_i^*$ and $\bfD_i^{**}$, including main effects and high-order interactions of $\bfX_i$. Deterministic functions of time, for example, basis spline functions of time in our EHR application, can be included in the covariates.
\begin{align}
\label{y}
	\bfY_i&=\bfD_i\bfbeta + \bfD_i^*\bfbeta^*_{C_i} + \bfD_i^{**}\mathbf{b}_i + \bfepsilon_i\\
	\nonumber \mathbf{b}_i&\sim N(\vec{0},\mathbf{\Sigma}_{r\times r})\mbox{, and } \bfepsilon_i \sim N(\vec{0},\sigma_{C_i}^2I_{T_i\times T_i}),
\end{align}
which is a linear mixed-effects model with a mixture of location and scale parameters varying across profiles. The profile-specific coefficients $\bfbeta^*_{C_i}$ can capture the differential trajectories of A1c measurements with time-varying covariates in $\bfD_i^*$. Model~\eqref{y} is a location and scale mixture model that is flexible enough to capture non-Gaussian distributions as shown in Figure~\ref{a1c_30}. Here $\mathbf{\Sigma}_{r\times r}$ is the covariance-variance matrix of the individual-specific random effects $\mathbf{b}_i$, which could be simplified as a diagonal matrix if the components of $\mathbf{b}_i$ are treated as independent or a scalar if $\mathbf{b}_i$ only represents random intercepts.

We assign weakly informative and conjugate prior distributions to the parameters~\citep{gelman2008}. The prior specification and full conditional posterior distributions as efficient Gibbs sampler are presented in Appendix~\ref{ps-ig}. After convergence, based on the posterior samples of the parameters, we can impute the missing A1c values and disseminate completed datasets.

\subsection{Joint profiling}

With MNAR, we jointly model the A1c values and missingness patterns given the covariates to obtain the joint profiling structure
\begin{align}
\label{nmar}
f(\bfY_i, \bfR_i|\bfX_{i0}, \bfX_i)=\sum_{l=1}^Lf(\bfY_i|\bfX_{i},C_i=l)f(\bfR_i|\bfX_{i},C_i=l)f(C_i=l|\bfX_{i0}),
\end{align}
where $\bfY_i$ includes both observed and missing measurements. The joint model has three components: the latent profile model, the longitudinal outcome model and the longitudinal response propensity model. 

This model captures the MNAR dependence between the outcome and missingness patterns that jointly determine the latent profiles. The response indicator is conditionally independent of the outcome given the latent profiles and covariates, i.e., conditional MAR: $R_{ij}\independent (Y_{ij},\bfX_{i0})|(C_i,\bfX_{ij})$. Conditional MAR makes models identifiable, yet flexible for dimension reduction and interpretation.

We consider a generalized linear mixed-effects model with a logit link for the longitudinal response indicator $R_{ij}$, conditional on latent classes and random effects $\mathbf{e}_{i}\sim N(0, \mathbf{E})$ with $\mathbf{E}$ as a covariance-variance matrix.
\begin{align}
\label{r}
\textrm{logit} \textrm{Pr}(R_{ij}=1|\bfX_{ij}, C_i, \mathbf{e}_{i}))=\bfD_{ij}\bfalpha+\bfD_{ij}^{*}\bfalpha^*_{C_i}+\bfD_{ij}^{**}\mathbf{e}_{i},
\end{align}
where the covariates $(\bfD_{ij}, \bfD_{ij}^{*}, \bfD_{ij}^{**})$ could overlap or be different from those selected in Model~\eqref{y} as subsets of the main effects and high-order interactions in $\bfX_{ij}$. 

\subsection{Summary}
\label{fullmodel}
The full model specification under marginal profiling is summarized below:
\begin{align}
\label{mar-all}
	\bfY_i&\sim N(\bfD_i\bfbeta + \bfD_i^*\bfbeta^*_{C_i} + \bfD_i^{**}\mathbf{b}_i, \sigma_{C_i}^2I_{T_i\times T_i})\\
	\nonumber C_i &\sim \textrm{Multinomial}(\pi_1(\bfX_{i0}), \dots, \pi_L(\bfX_{i0}))\\
	\nonumber &\pi_l(\bfX_{i0})=\frac{\exp(\bfX_{i0}^T\bfeta_l)}{\sum_{k=1}^L\exp(\bfX_{i0}^T\bfeta_k)}\\
	%\nonumber w_{il} &\sim \textrm{PG}(1,r_{il})\mbox{, where } r_{il}=\bfX_{i0}^T\bfeta_l-\log[\sum_{k\neq l}\exp(\bfX_{i0}^T\bfeta_k)]\\
	\nonumber \mathbf{b}_i&\sim N(\vec{0},\mathbf{\Sigma}_{r\times r})\mbox{, }\beta\sim N(0,\mathbf{\Sigma}_{\beta})\mbox{, }\mathbf{\Sigma}_{r \times r} \sim \textrm{Inverse-Wishart}(\nu_b,\mathbf{\Sigma}_b)\\
	\nonumber \sigma_l^{2}&\sim \textrm{Inverse-Gamma}(a,b)\mbox{, }\eta_l\sim N(b_l, B_l)\mbox{, for } l=1,\dots, L.
\end{align}

The full model specification under joint profiling is summarized below:
\begin{align}
\label{joint-all}
	\bfY_i&\sim N(\bfD_i\bfbeta + \bfD_i^*\bfbeta^*_{C_i} + \bfD_i^{**}\mathbf{b}_i, \sigma_{C_i}^2I_{T_i\times T_i})\\
	\nonumber R_{ij} &\sim \textrm{Bernoulli}(\textrm{Pr}(R_{ij}=1))\\
	\nonumber  &\textrm{Inverse-logit}(\textrm{Pr}(R_{ij}=1))=\bfD_{ij}\bfalpha+\bfD_{ij}^{*}\bfalpha^*_{C_i}+\bfD_{ij}^{**}\mathbf{e}_{i}\\
	\nonumber C_i &\sim \textrm{Multinomial}(\pi_1(\bfX_{i0}), \dots, \pi_L(\bfX_{i0}))\\
	\nonumber &\pi_l(\bfX_{i0})=\frac{\exp(\bfX_{i0}^T\bfeta_l)}{\sum_{k=1}^L\exp(\bfX_{i0}^T\bfeta_k)}\\
	%\nonumber w_{il} &\sim \textrm{PG}(1,r_{il})\mbox{, where } r_{il}=\bfX_{i0}^T\bfeta_l-\log[\sum_{k\neq l}\exp(\bfX_{i0}^T\bfeta_k)]\\
	\nonumber \mathbf{b}_i&\sim N(\vec{0},\mathbf{\Sigma}_{r\times r})\mbox{, } \beta\sim N(0,\mathbf{\Sigma}_{\beta})\mbox{, }\mathbf{\Sigma}_{r\times r} \sim \textrm{Inverse-Wishart}(\nu_b,\mathbf{\Sigma}_b)\\
	\nonumber \alpha_l &\sim N(0,\mathbf{\Sigma}_{\alpha})\mbox{, }\eta_l\sim N(b_l, B_l)\mbox{, }\sigma_l^{2}\sim \textrm{Inverse-Gamma}(a,b)\mbox{, }\\
	\nonumber \nu&\sim N(0,\mathbf{\Sigma}_{\nu})\mbox{, }\gamma_l \sim N(0, \mathbf{\Sigma}_{\gamma})\mbox{, }\mathbf{e}_{i}\sim N(0, \mathbf{E}) \mbox{, }\mathbf{E} \sim \textrm{Inverse-Wishart}(\nu_e, \mathbf{\Sigma}_e).
\end{align}

We set the hyper-parameters equal to 1 and specify an independent covariance matrix structure. Our inference is not sensitive to the specification of hyper-parameter values (e.g., 0.1, 0.01) under noninformative or weakly informative prior settings, as expected given the large sample sizes in the application. The PG variables $w_{il}$'s are introduced for posterior updating with $\bfeta$. We develop the Gibbs sampler with two sets of introduced P\'{o}lya-Gamma distributed variables for posterior computation with models~\eqref{pi}, \eqref{y} and \eqref{r} under weakly informative and conjugate prior distributions. The imputation of missing A1c values is nested in the iterative process, unlike marginal profiling, where parameter estimation and data augmentation are implemented simultaneously. Computational details are presented in Appendix~\ref{ps-nonig}.

For ease of interpretation of the potential profiling structure, we fix the number of latent patterns $L$ to be a small integer, chosen with the aid of diagnostic tools for model selection described in Section~\ref{diagnostics}. The model can be extended to allow nonparametric Bayesian modeling, such as the dependent probit stick-breaking process~\citep{abel2011}. However, discrete covariates with low prevalence tend to cause separation problems with a large number of latent profiles, as in our application study. 

The BPMI in~\eqref{mar-all} and~\eqref{joint-all} uses mixture models to flexibly capture irregular distributions, incorporate time trends, and latent profiles allow us to jointly model the outcomes and nonignorable missingness patterns.

%%%%%%%%%%%%%%%%%%%%%%%%%%%%%%%%%%%%%%
\section{Simulation}
\label{simulation}

We evaluate the imputation quality and MI inferential validity in repeated sampling. We simulate unbalanced longitudinal data, and the outcome is subject to intermittent missing values. For each sample, we generate $n=500$ subjects with varying lengths of follow-up periods $T_i$'s randomly selected from $\{1,\dots, 10\}$ and time-invariant covariates $D_i$ with four binary indicators simulated from Bernoulli's distributions with probabilities 0.2, 0.3, 0.4, and 0.5, respectively. The time-varying covariate is the time after the first collection, where we use $t_{ij}$ denoting the time points of collection divided by 4 to mimic the quarter structure in the EHR data. We consider different outcome generation models and different missingness mechanisms, as we now describe.

In Case 1, we groups the 500 subjects into $L=3$ classes with allocation probabilities $\textrm{exp}(D_i\eta_l)/(1+\textrm{exp}(D_i\eta_l))$, where $\eta_1=\vec{0}$, $\eta_l=0.5 * (1:5) - 0.5 *l$, for $l=2, 3$. The outcome $Y$ is simulated from a 3-component finite mixture of normal distributions with mean $\mu_{ij}=b_i + D_i\beta_0 + t_{ij}\beta_{1l}+t_{ij}^2\beta_{2l}$ and standard deviation (sd) $\sigma_{l}$, for $l=1,2,3$, respectively. Here we set $\beta_0=(2, 2, 2, 2, 2), \beta_{11}=0, \beta_{12}=-2, \beta_{13}=2, \beta_{21}=0, \beta_{22}=1, \beta_{13}=1, \sigma_1=0.1, \sigma_2=1$, and $\sigma_3=3$. We assume the intermittent missingness is MAR with the response probability $\textrm{Pr}(R_{ij}=1)=\textrm{exp}(D_i\alpha_0 + t_{ij})/(1+\textrm{exp}(D_i\alpha_0 + t_{ij}))$ for $j=2,\dots, T_i, i=1,\dots n$, where $\alpha_0=(-2, -1, 0, 1, 2)$. The missingness mechanism depends on the main effects of the covariates and time. The missingness proportion is around $38\%$. The data generation and missingness mechanism in Case 1 are consistent with the outcome models and identification assumptions under marginal profiling, denoted as Mixture--Main.

In Case 2, the outcome is simulated from a normal distribution with mean $\mu_{ij}=b_i + D_i t_{ij} \beta_0 + t_{ij}\beta_{1}$ and sd=1, where the interaction term between the time-invariant covariates and time are included with $\beta_0=(2, 2, 2, 2, 2)$ and $\beta_1=1$. The model assumes that the linear trends vary across individuals. The missingness mechanism is the same as that in Case 1. We denote the data generation and missingness mechanism in Case 2 as Interaction--Main, which differs from the outcome model assumptions under BPMI.

In Case 3, the outcome model has the same specification as that in Case 2, with interaction terms. The missingness also depends on the interaction terms between the time-invariant covariates and time: $\textrm{Pr}(R_{ij}=1)=\textrm{exp}(e_{i} + D_i\alpha_0 + t_{ij}\alpha_{1}+D_it_{ij}\alpha_{2})/(1+\textrm{exp}(e_{i} + D_i\alpha_0 + t_{ij}\alpha_{1}+D_it_{ij}\alpha_{2}))$ for $j=2,\dots, T_i, i=1,\dots n$, where $\alpha_0=(1, 1, 1, 1, 1), \alpha_{1}=-0.5, \alpha_2=(-2, -1, 0, 1, 2)$, and $e_{ij} \sim N(0,1)$, with resulted missingness around 20\%. We denote the data generation and missingness mechanism in Case 4 as Interaction--Interaction, which differs from assumptions for both the outcome and missingness mechanism under BPMI.

In Case 4, the outcome model is the same as that in Case 1. We assume that the missingness depends on the group structure, and it is conditionally MAR but marginally MNAR. The response probability is $\textrm{Pr}(R_{ij}=1)=\textrm{exp}(e_{i} + D_i\alpha_0 + t_{ij}\alpha_{1l}+t_{ij}^2\alpha_{2l})/(1+\textrm{exp}(e_{i} + D_i\alpha_0 + t_{ij}\alpha_{1l}+t_{ij}^2\alpha_{2l}))$ for $j=2,\dots, T_i, i=1,\dots n$, where $\alpha_0=(1, 1, 1, 1, 1), \alpha_{11}=0, \alpha_{12}=-2, \alpha_{13}=2, \alpha_{21}=0, \alpha_{22}=-4, \alpha_{23}=4$, and $e_{ij} \sim N(0,1)$, resulting in around 35\% of missing values. The scenario in Case 3 is similar to that under joint profiling, denoted as Mixture--Mixture.

In Case 5, we violate the conditional MAR assumption by specifying the response propensity still depends on the outcome given the group structure. The remaining specifications are the same as those in Case 4. This case is denoted as Mixture--Outcome.

We implement the marginal profiling approach (BPMI-MAR), the joint profiling approach (BPMI-MNAR), with the popular imputation methods MICE and multilevel imputation MICE-2l, to generate 10 imputed datasets each. Both MICE and MICE-2l assume MAR. MICE treats measures at the 10 time points as 10 different variables, performs fully conditional chained imputation with linear regression models and fills in missing values with draws from observed cases based on predictive mean matching by default. This accounts flexibly for the correlation between repeated measures but ignores any pattern across time. MICE generates balanced data, where measures at all time occasions are filled in even though some are not eligible in the intrinsic data structure. MICE-2l fits a multilevel model for the outcome with a random intercept and a random slope for time. MICE-2l fits a linear time trend, and assumes normality of all random components. Hence, BPMI differs from the other two approaches in being able to handle MNAR, having more flexible distributional assumptions, modeling the time trend in a flexible manner and in handling covariates in an implicit manner with potential interaction terms rather than by an explicit inclusion of main effects.

\begin{figure}
  \centering
  \vspace{-1cm}
  \begin{tabular}{cccc}
\includegraphics[width=0.31\textwidth,height=1.3in]{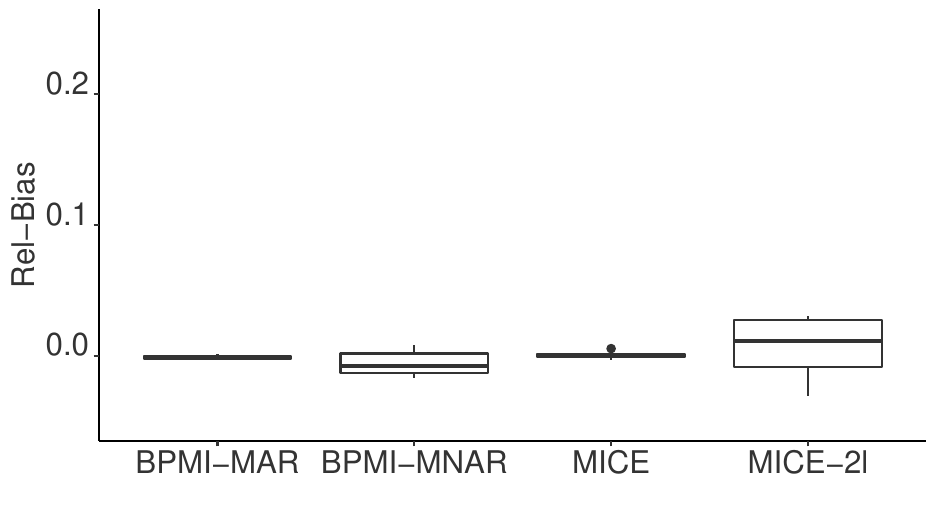} &\includegraphics[width=0.31\textwidth,height=1.3in]{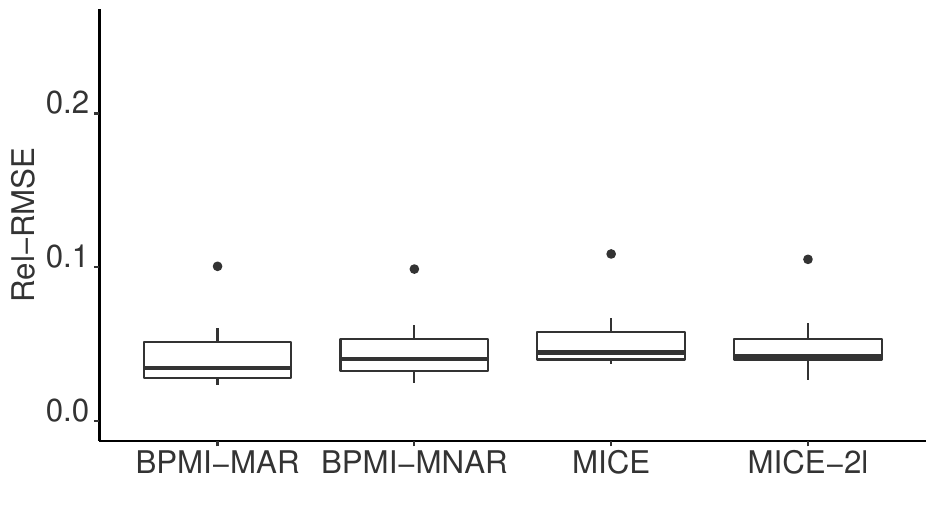} &\includegraphics[width=0.31\textwidth,height=1.3in]{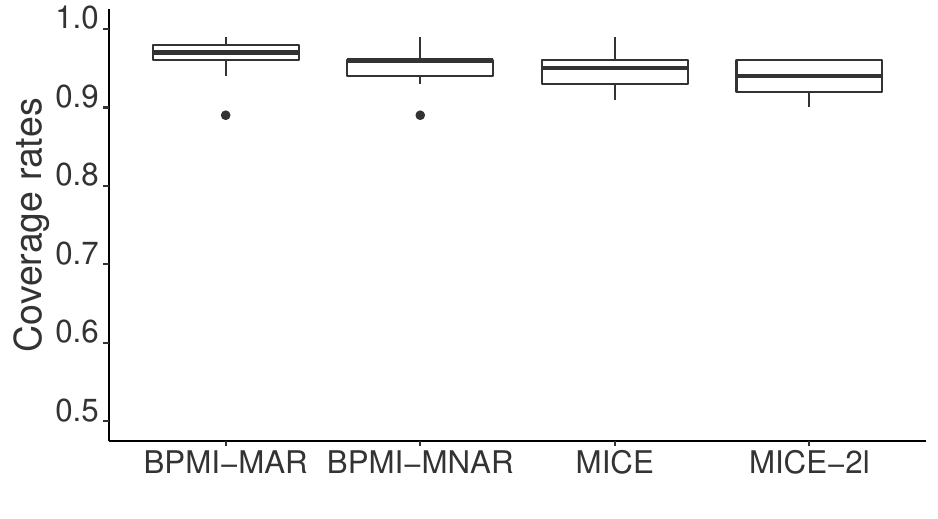} \\
\vspace{-0.8cm}
\end{tabular}
\subfloat[\label{case1} Case 1: Mixture--Main.]{\hspace{.95\textwidth}}\\
  \begin{tabular}{cccc}
\includegraphics[width=0.31\textwidth,height=1.3in]{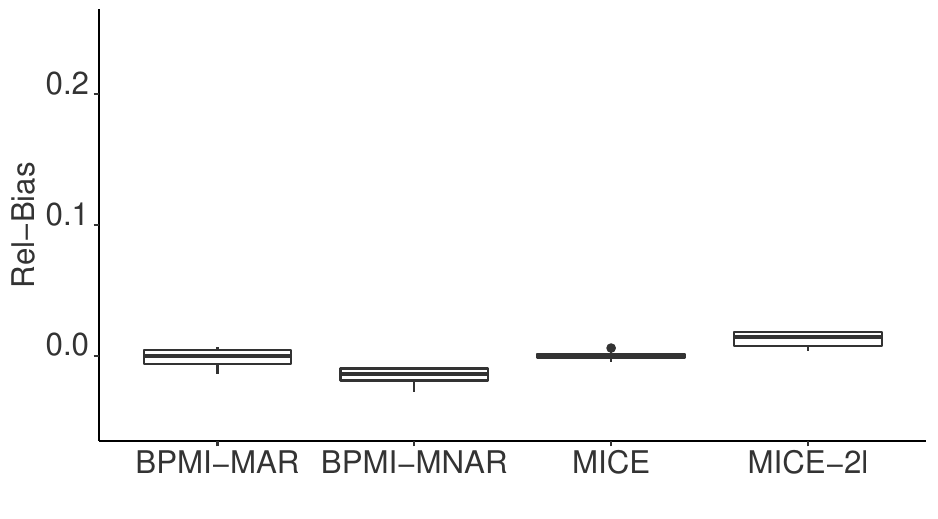} &\includegraphics[width=0.31\textwidth,height=1.3in]{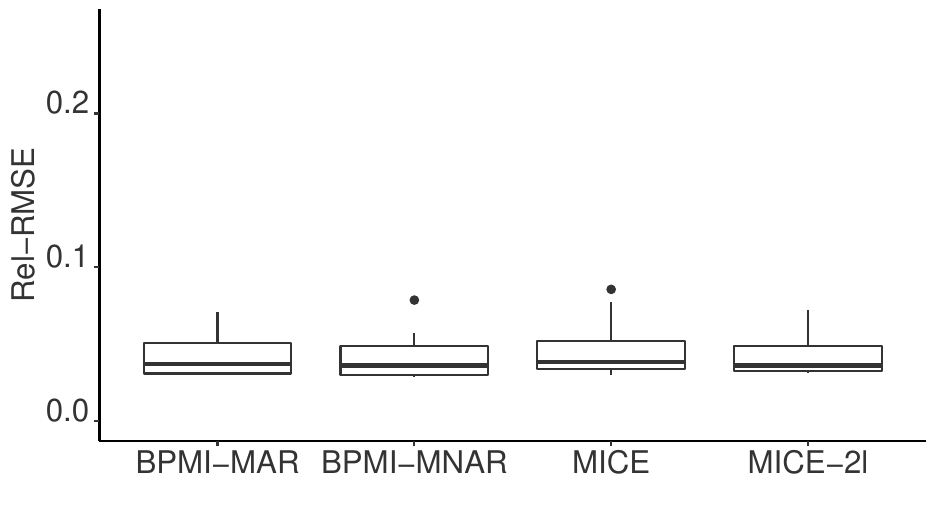} &\includegraphics[width=0.31\textwidth,height=1.3in]{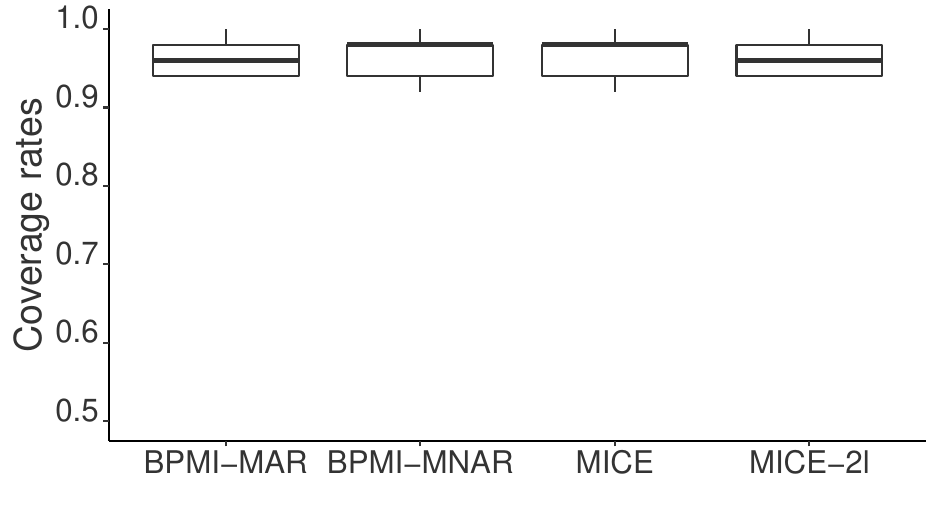} \\
\vspace{-0.8cm}
\end{tabular}
\subfloat[\label{case2} Case 2: Interaction--Main.]{\hspace{.95\textwidth}}\\
  \begin{tabular}{cccc}
\includegraphics[width=0.31\textwidth,height=1.3in]{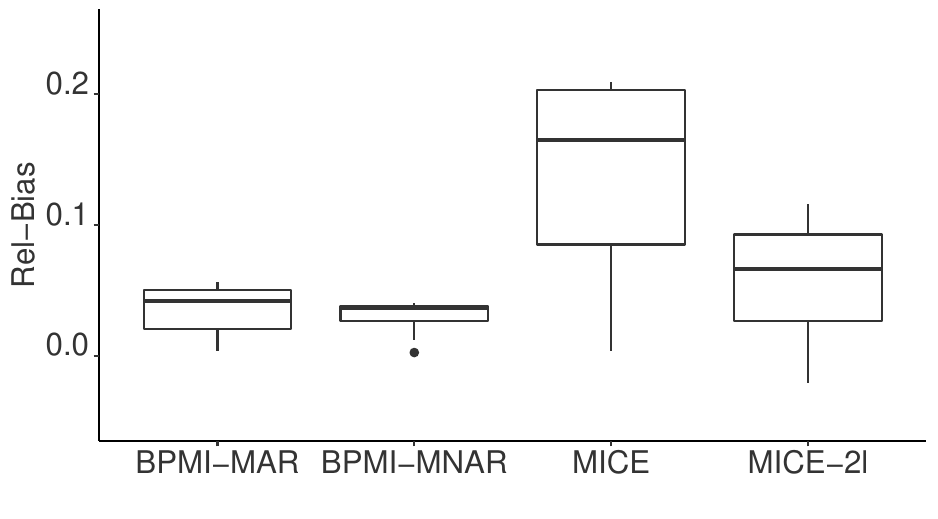} &\includegraphics[width=0.31\textwidth,height=1.3in]{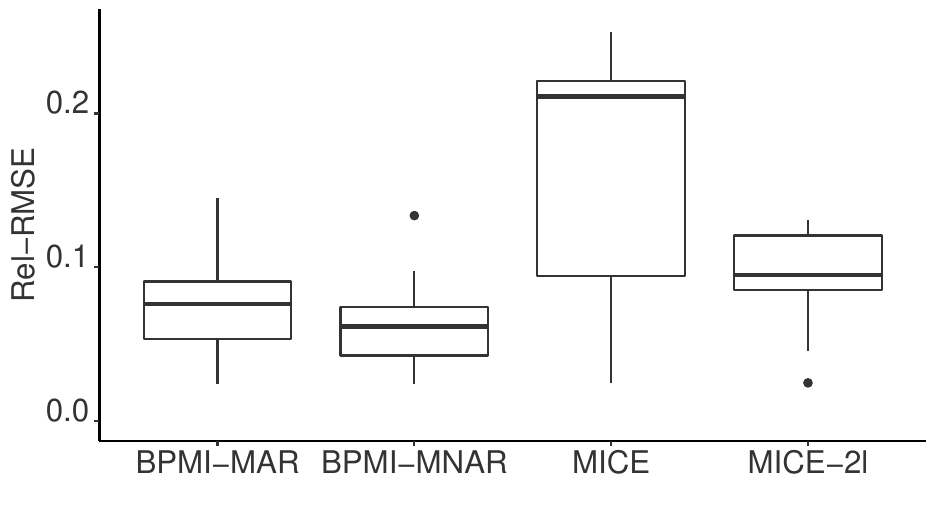} &\includegraphics[width=0.31\textwidth,height=1.3in]{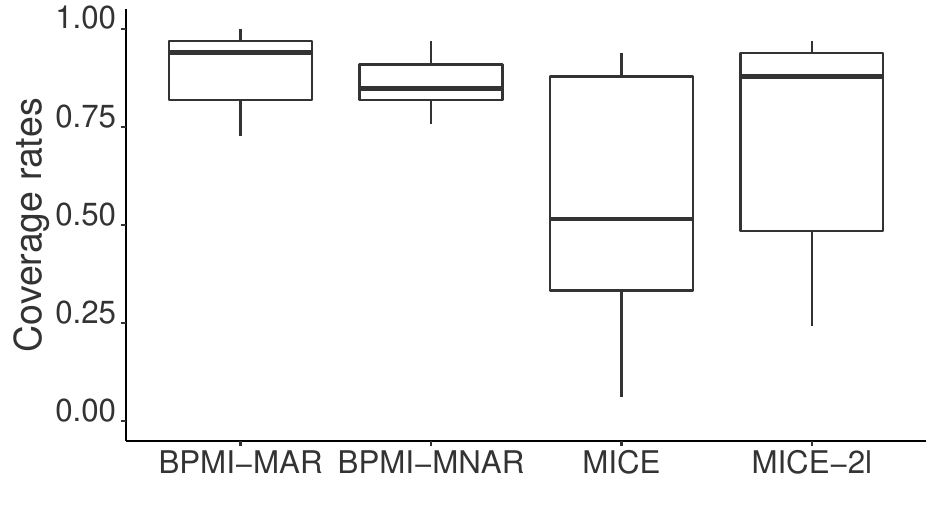} \\
\vspace{-0.8cm}
\end{tabular}
\subfloat[\label{case3} Case 3: Interaction--Interaction.]{\hspace{.95\textwidth}}\\
  \begin{tabular}{cccc}
\includegraphics[width=0.31\textwidth,height=1.3in]{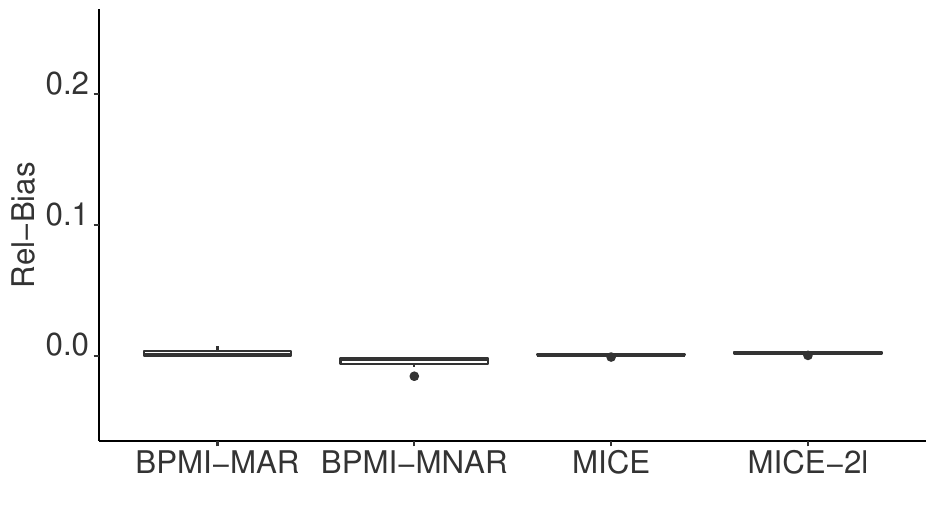} &\includegraphics[width=0.31\textwidth,height=1.3in]{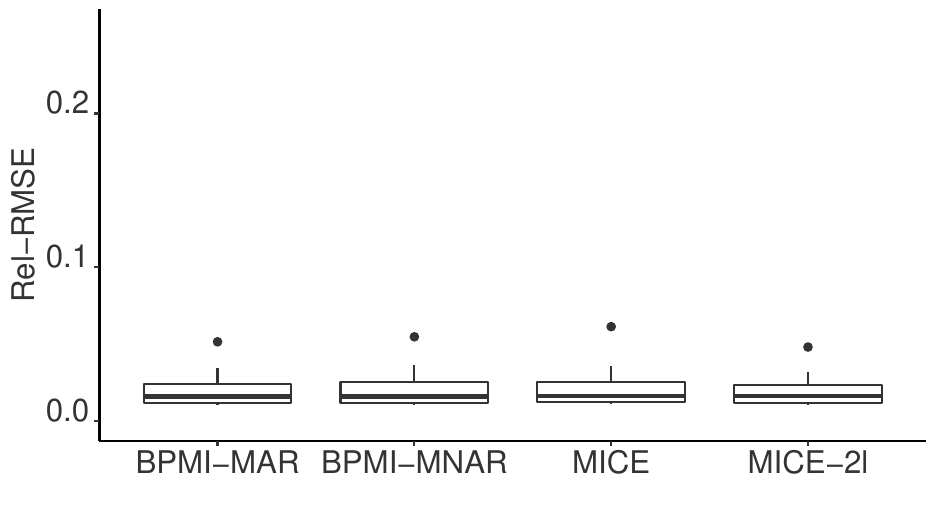} &\includegraphics[width=0.31\textwidth,height=1.3in]{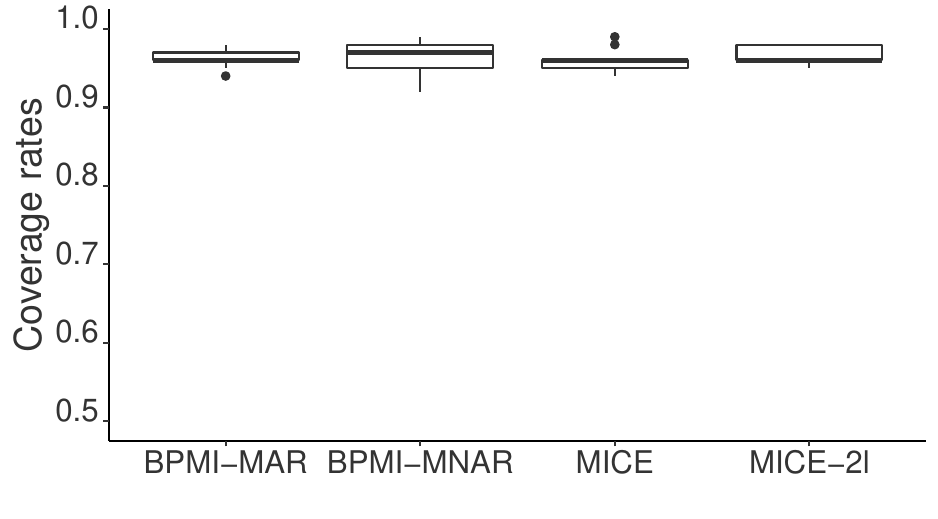} \\
\vspace{-0.8cm}
\end{tabular}
\subfloat[\label{case3} Case 4: Mixture--Mixture.]{\hspace{.95\textwidth}}\\
  \begin{tabular}{cccc}
\includegraphics[width=0.31\textwidth,height=1.3in]{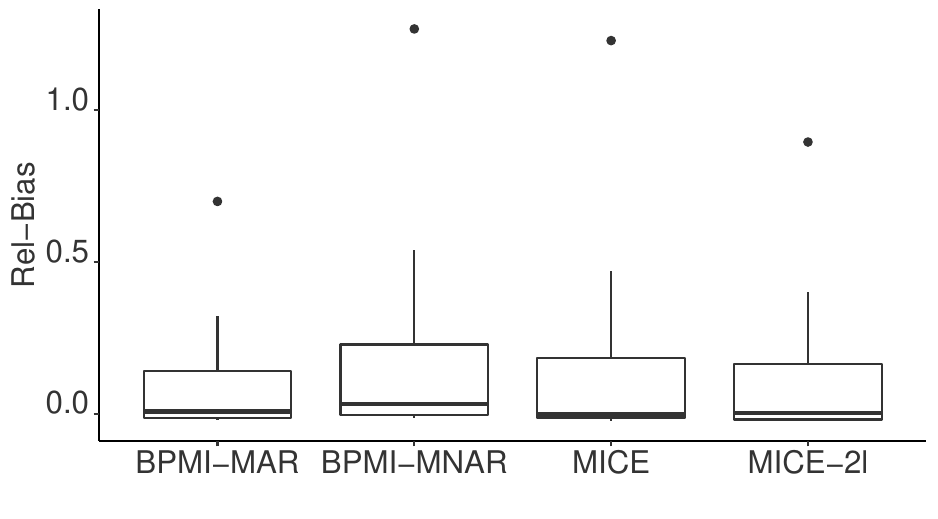} &\includegraphics[width=0.31\textwidth,height=1.3in]{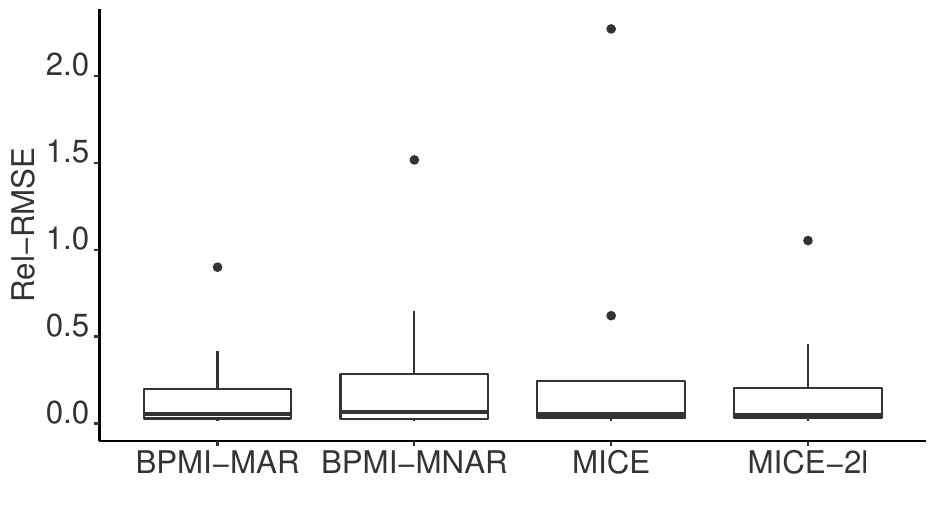} &\includegraphics[width=0.31\textwidth,height=1.3in]{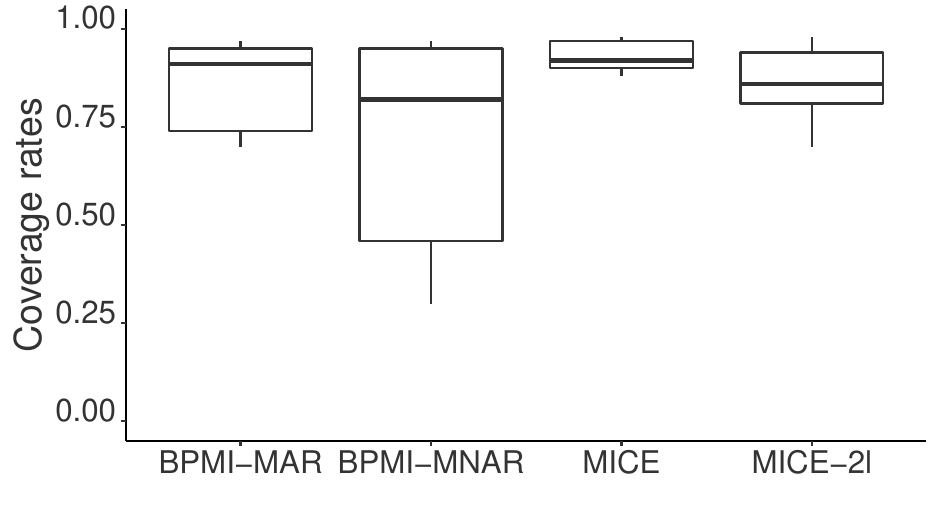} \\
\vspace{-0.8cm}
\end{tabular}
\subfloat[\label{case3} Case 5: Mixture--Outcome.]{\hspace{.95\textwidth}}\\
 \caption{Performance of MI inferences on the average outcome values at 9 follow-up occasions with 10 imputed datasets comparing different approaches: marginal profiling (BPMI-MAR), joint profiling (BPMI-MNAR), sequential imputation MICE and multilevel imputation MICE-2l.}
  \label{sim-fig}
\end{figure}

We run BPMI-MAR in the exact specification in~\eqref{mar-all} for 3000 iterations and BPMI-MNAR in~\eqref{joint-all} for 2000 iterations with $L=3$. The time-varying covariates in BPMI include the basis spline functions of time used in Section~\ref{application} and given in Appendix~\ref{spline}. Our quantities of interest are the average values of the outcome in the follow-up period, across $j=2,\dots,10$. We perform repeated sampling for 100 times and stack the generated samples treated as the population. We combine the MI inferences and calculate average values of the bias, root mean squared error (RMSE) and nominal coverage rate of  95\% confidence intervals for each of these four approaches. We divide the bias and RMSE by the true value and present the relative bias (Rel--Bias) and relative RMSE (Rel--RMSE), and coverage rates in Figure~\ref{sim-fig}.

The marginal profiling approach is the most robust and efficient method across the five cases. In Case 1, the sequential imputation MICE generates unbiased estimates but with larger RMSEs and smaller coverage rates than the BPMI-MAR. The performance of joint profiling BPMI-MNAR is competitive, with relatively larger bias and RMSE but negligible differences. The coverage rates of multilevel imputation models MICE-2l are generally lower than 0.95 even with large RMSEs. 

Case 2 shows that BPMI can recover the interaction effects through latent profiling when the covariates enter through the allocation probabilities but are excluded in the outcome modeling and imputation. This is also confirmed in Case 3, when interaction terms are present in the response propensity model. In Case 4, with marginal MNAR and conditional MAR, MICE and MICE-2l fail to yield valid inferences as both methods assume MAR. The BPMI approaches outperform with small bias, RMSEs and reasonable coverage rates. The outputs of joint profiling BPMI-MNAR did not show improvement over those under marginal profiling. This could be due to the Monte Carlo error with not enough number of repeated samples or the computation problems of BPMI-MNAR that requires more iterations to converge. Both BPMI and MICE generate large biases and RMSEs in Case 5, when the missing data mechanism assumption is violated.

%%%%%%%%%%%%%%%%%%%%%%%%%%%%%%%%%%%%%
\section{Imputation and Inference with EHRs}
\label{application}

In a two-step MI process, we first apply BPMI to impute missing A1c values in the EHRs and secondly make inference on how A1c relates to acute diabetes outcomes. The observed A1c values range from 3.3\% to 17.3\% with mean 7.1\% and are right-skewed with multiple modes. We use the location and scale mixture distribution as a flexible strategy to capture the irregular distribution. In the profile allocation model~\eqref{pi}, the time-invariant covariates $\bfX_{i0}$ are the baseline characteristics listed in Table~\ref{predictor}. In the A1c outcome and response propensity models~\eqref{y} and~\eqref{r}, the covariates in $\bfD_{ij}$ with fixed effects include the constant ones and the time-varying variables, age and the count of physician visits. The covariates in $\bfD^*_{ij}$ represent the functions of time in years with slopes changing across classes. We use the basis spline functions of time with $(1\%, 15\%, 20\%, 50\%, 75\%, 90\%)$ quantiles as six knots and cubic terms as a piecewise polynomial regression, and the nine spline functions used as covariates in $\bfD^*_{ij}$ with profile-specific coefficients as shown in Appendix~\ref{spline}. The spline functions are chosen based on model fitting criteria and can handle complex shapes to create smooth curves. Details on diagnostics of model specifications are provided in Section~\ref{diagnostics}. Random intercepts are introduced by $\bfD^{**}_{ij}$.

To set up the initial values of the Markov chain Monte Carlo (MCMC) updating, we fit an ordinary linear regression and use the coefficient estimates as initial values for $\bfbeta$ and $\bfbeta^*_{l}$'s whose starting points are the same across classes. The initial values of $\bfeta$'s are $\vec{0}$, representing an initial equal probability of class allocation. The scale parameters start at 0.1, and the random effects are initially drawn from normal distributions with mean 0 and the initial scales.

We implement posterior computation for both marginal and joint profiling. The MCMC algorithms efficiently achieve convergence with Gibbs samplers as diagnosed by trace-plots of the posterior samples, where 15000 iterations under marginal profiling and 5000 iterations under joint profiling take $\sim$5 hours to finish each with R codes run on a standard laptop (3.5 GHz Dual-Core Intel Core i7, 16 GB Memory), and the computational speed can be substantively improved with R--C++ interface and parallel computing resources. We run the MCMC chains long enough to obtain 100 multiply imputed datasets and 500 replicated datasets to check if imputation is able to predict observed outcomes. For classification, we follow~\cite{goodman07} with a hard partitioning and keep one random assignment based on a random draw from the component-specific probabilities across iterations. Our analysis shows that the results are not sensitive to the classification rules.

\subsection{Model diagnostics and comparison}
\label{diagnostics}
We use the Bayesian information criterion (BIC) and the log-pseudo marginal likelihood (LPML) to select the number of classes, $L$. For BIC, we use the posterior mean estimates of related parameters and obtain the likelihood conditional on the class allocation. LPML is the sum of logarithms of conditional predictive ordinates and estimated using posterior samples of parameters, $\theta^{(t)},t=1,\dots,T$, and LPML=$[1/T\sum_{t=1}^T1/f(Y_i|\theta^{(t)})]^{-1}$~\citep{gelfand94}. This is an approximation of the leave-one-out cross-validation using importance sampling. We use $f(Y_i|\theta^{(t)})$ for the marginal profiling and $f(Y_i,R_i|\theta^{(t)})$ for the joint profiling. 

\begin{table}
\caption{Comparison of models with different number of profiles ($L$).}
\label{lsm-L}
\centering
\begin{tabular}{lcc|lcc}
\hline
\multicolumn{3}{c}{Marginal profiling} & \multicolumn{3}{c}{Joint profiling}\\
\hline
L&BIC&LPML&L&BIC&LPML\\
\hline
2&126857&-106129&2&{\bf 323529}&-192506\\
3&{\bf 124145}&{\bf -94305}&3&329037&{\bf -174286}\\
4&130872&-97579&4&334345&-180697\\
5&136263&-97315&&\\
6&137082&-97397&&\\
\hline
\end{tabular}
\end{table}

We restrict $L$ to be small to avoid computational problems due to the separation of the low-frequency predictors, such as dementia and renal failure. Table~\ref{lsm-L} presents the BIC and LPML values for models with different $L$ values. The model with the smallest BIC and largest LPML will be favored. We see that the model with $L=3$ classes presents a reasonable choice both for marginal and joint profiling.

To assess whether imputations are plausible, we perform a posterior predictive check and generate replicated datasets that are predicted values of observations based on the posterior estimates of model parameters after convergence~\citep{meng1994}. Let $\{R^{(1)},\dots,R^{(T^0)}\}$ be the collection of the $T^0$ replicated data sets. We then compare statistics of interest in each replicated dataset to those in the observed dataset $D$. Suppose that $S$ is some statistic of interest, let $S_{R^{(t)}}$ and $S_D$ be the values of $S$ computed from $R^{(t)}$ and $D$, respectively. The quantities $S$ include the mean, 2.5\% percentile and 97.5\% percentile of A1c levels for every patient and for every quarter. For the patient-level summaries, we look at the two-sided posterior predictive probabilities as diagnostic tools defined as
\[
\textrm{ppp}=\frac{2}{T^0}* \textrm{min}(\sum_{t=1}^{T^0}I(S_{R^{(t)}}-S_D>0), \sum_{t=1}^{T^0}I(S_{R^{(t)}}-S_D<0)).
\]
When the value ppp is small, for example, less than 5\%, this suggests that the replicated data sets systematically differ from the observed data set, with respect to that statistic. With larger ppp values, the evidence does not contradict that the imputation model preserves the observed characteristics of that statistic~\citep{he2010,blpm-si:15}. For the 7372 patient-level A1c summaries, the marginal profiling model yields 52 ppp values that are below 1\% for the average, and the numbers of below 1\% ppp values for the 2.5\% percentile and the 97.5\% percentile are 238 and 375. The number of ppp values that are below 1\% for these three statistics under joint profiling is 66, 464 and 678, respectively. Based on the low proportions of small ppp values, the posterior predictive check does not indicate lack of model fit in recovering the trajectories. However, the performance of joint profiling is inferior to marginal profiling in preserving the observed mean values across patients.

\begin{figure}
\begin{tabular}{cc}
\includegraphics[width=0.5\textwidth,height=3in]{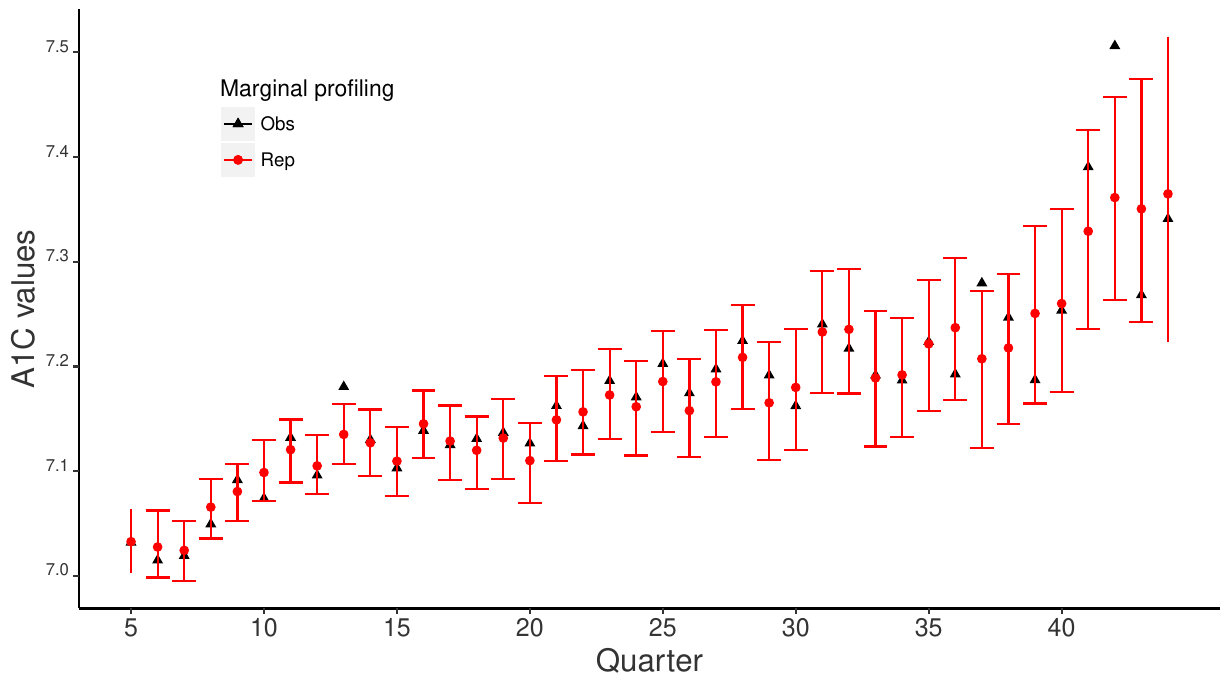}&
\includegraphics[width=0.5\textwidth,height=3in]{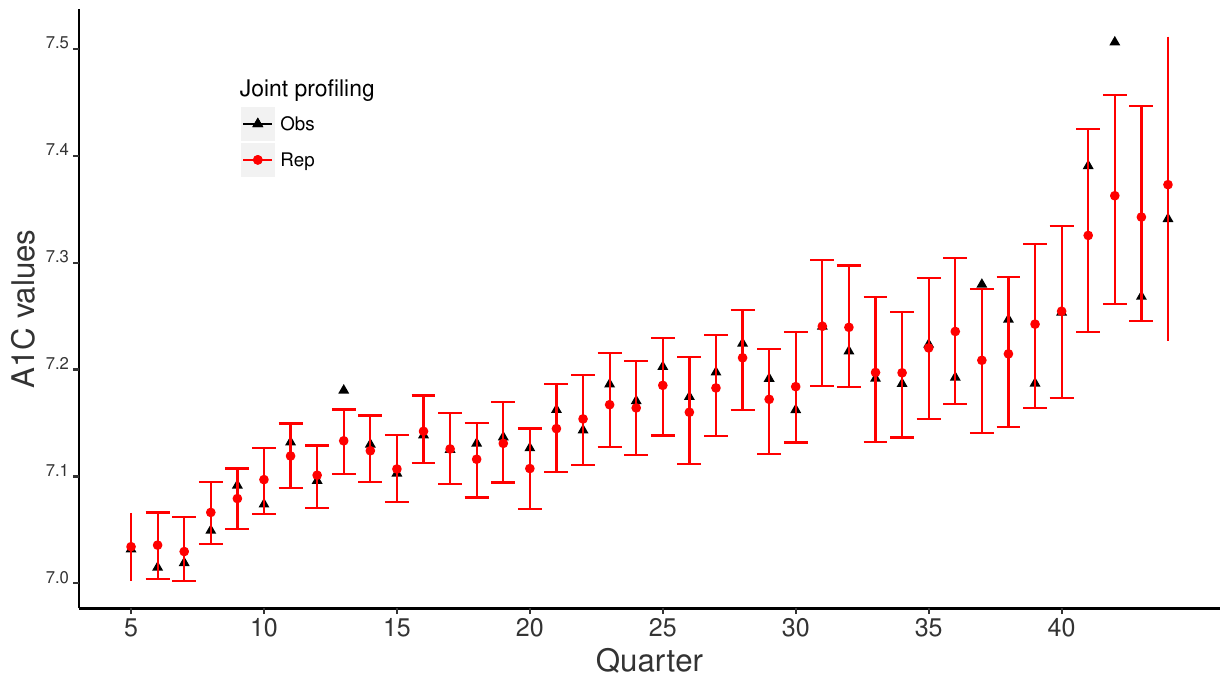}\\
\end{tabular}
\caption{Posterior predictive check on quarter-level summaries for marginal and joint profiling approaches. The black triangles represent the means of observed data, the red dots are the posterior predictive mean estimates, and the red error bars are the 95\% credible intervals of the predictive means.}
\label{mar_nmar}
\end{figure}

For the quarter-level A1c summaries, we compare the predictions with the observed values under the two profiling approaches. The 95\% predictive credible intervals greatly overlap with the empirical 95\% confidence intervals (CIs). Figure~\ref{mar_nmar} shows that the posterior predictive mean estimates are generally close to the observed mean values, except for two observations, and both approaches perform competitively and similarly. We check the parameter estimates in Model~\ref{joint}, and find that the coefficients do not change across profiles. This explains the similar performance of joint profiling and marginal profiling.

\begin{figure}
\begin{tabular}{c}
\includegraphics[width=0.99\textwidth]{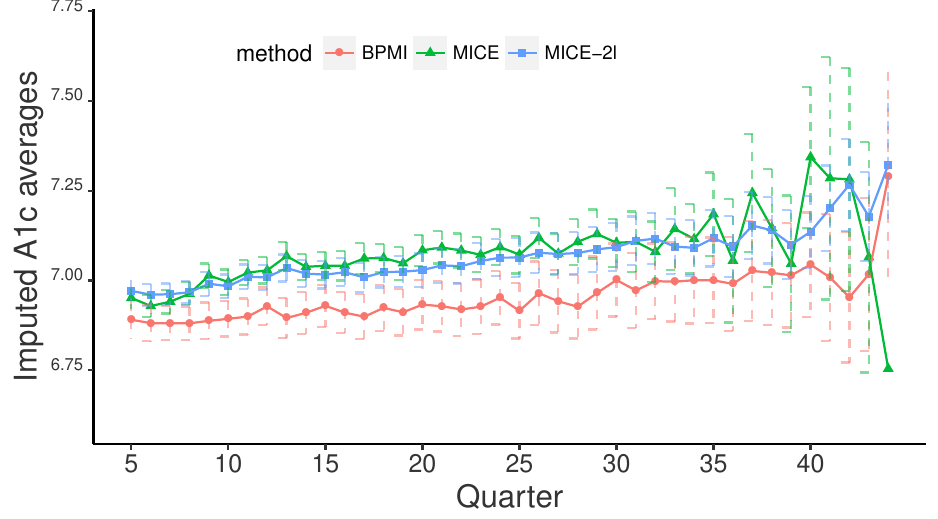}
\end{tabular}
\caption{Comparison of the average of imputed A1c values and 95\% confidence intervals obtained by BPMI, MICE and MICE-2l across follow-up quarters. The wide CI of the MICE output for the last quarter (6.11, 7.40) is omitted. }
\label{imp_bpmi_mice}
\end{figure}

We also compare BPMI with MICE and MICE-2l, by examining the summaries of the imputed A1c values. Figure~\ref{imp_bpmi_mice} depicts the average values and 95\% confidence intervals of A1c measurements in the follow-up period from the 5th to the 44th quarter. We do not have a gold standard for the missing values and aim to check which method can recover the data structure. With the MAR assumption for MICE and BPMI under marginal profiling, the time dependence structure should be the same between the imputed data and observed data. Comparing the trajectories of the imputed data to those of the observed data in Figure~\ref{mar_nmar}, marginal profiling under BPMI yields imputation that presents the patterns most similar to the observed data, showing the trend of BPMI first flat and then increasing. The failure of MICE to impose a time structure results in wiggly averages and the last measurement having the lowest value. In addition, the need for MICE to impute occasions post drop-out or death may have led to inconsistent means in the later quarters. The multilevel imputation MICE-2l smooths the variation across time and presents a linear trend, which is different from the observed trends in Figure~\ref{mar_nmar}. Both MICE approaches tend to generate larger imputed values than those under BPMI, potentially due to the ability of BPMI to better replicate a multimodal skewed distribution of A1c values. 

In fact, BPMI imputes a substantially higher proportion of A1c values in the 6-7\% range and fewer in the 7-8\% range or below 6\% than do MICE or MICE-2l. Based on a randomly selected single imputed dataset, 18.3\% of the imputed values from BLPM are below 6\%, 44\% values are between 6\% and 7\%, and 23\% values are in the 7-8\% range. However, the imputed values from MICE have 20\% below 6\%, 37\% between 6\% and 7\%, and 27\% in the 7-8\% range, and the percentages from MICE-2l are 21\%, 36\% and 28\%, respectively. Having more imputed data in the below 7\% range resonates clinically with the fact that patients in that range have met ADA goals and may be seen as needing less frequent A1c testing.

%\begin{table}[ht]
%\centering
%\caption{Percentages}
%\begin{tabular}{clll}
%  \hline
% A1c levels & MICE & MICE-2L &BLPM \\ 
%  \hline
% $<$6\% & 0.203 & 0.207 &0.183 \\ 
% 6$\sim$ 6.5\% & 0.177 & 0.173 &0.223 \\ 
% 6.5$\sim$7\% & 0.194 & 0.189 &0.215 \\ 
%7$\sim$7.5\% & 0.163 & 0.164 &0.147 \\ 
%7.5$\sim$8\% & 0.11 & 0.111 &0.089 \\ 
% 8$\sim$9\% & 0.103 & 0.106& 0.087 \\ 
% 9\%+ & 0.05 & 0.051 &0.056 \\ 
%   \hline
%\end{tabular}
%\end{table}

\subsection{Inference and profiling}
 
After MI, we analyze the completed datasets to evaluate the association of A1c with acute health event incidence and depict the different trajectories of A1c values across time indicated by the profiling structure.

\subsubsection{Association of A1c with acute adverse events}
We aim to address whether using imputed data affects inferences on how glycemic control is associated with acute diabetes outcomes. Since marginal profiling and joint profiling present similar results, we use the results under marginal profiling as an illustration. It should be noted that the MAR assumption here applies to missing A1c. The complete case analysis (CCA) applies to the occasions with collected A1c values and causes exclusion of the outcomes available at the occasions where A1c values are missing. Hence, missingness in the outcome is induced by missing A1c and may be MAR or MNAR, but imputing A1c restores all the outcomes. 

\begin{figure}
\begin{tabular}{c}
\includegraphics[width=0.85\textwidth]{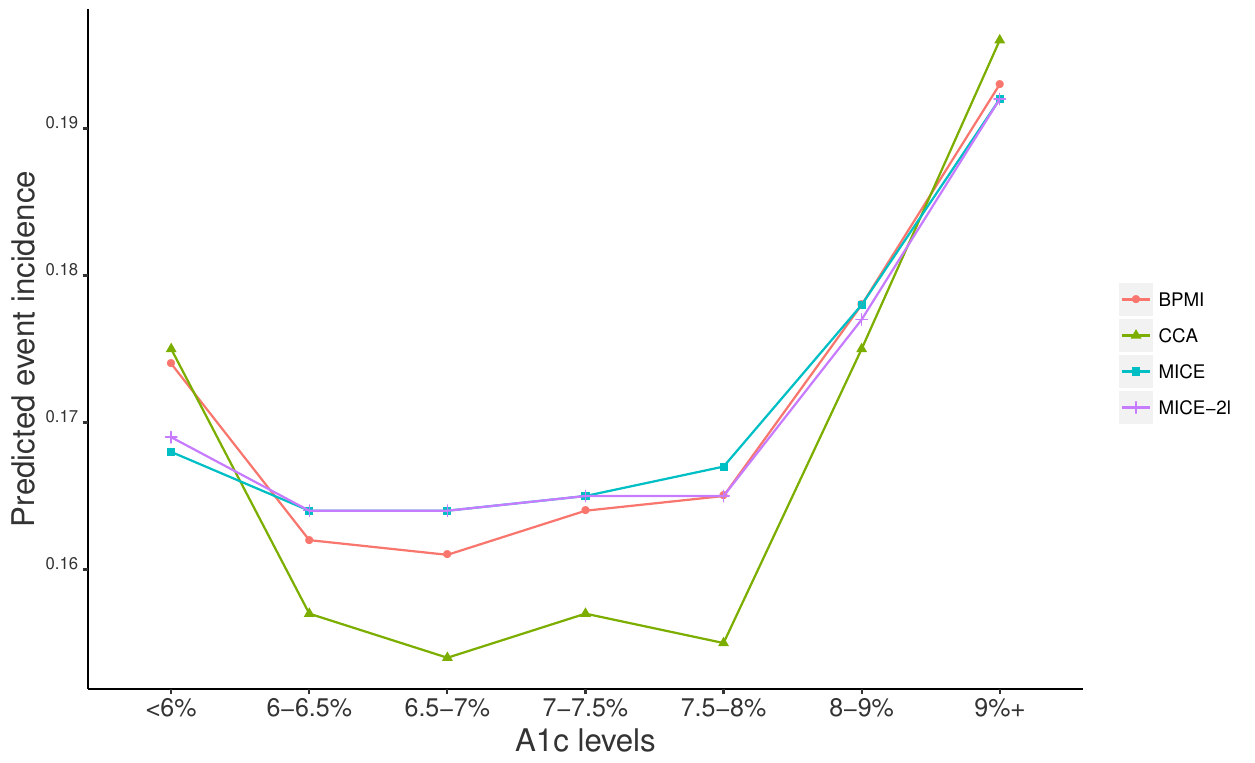}
\end{tabular}
\caption{Comparison between MI inferences and complete case analysis (CCA) in predictive acute health event incidence risk. }
\label{risk-pred}
\end{figure}

The overall incidence of any acute health events is 16\% across the patient-quarters. We predict this incidence by a seven-category discrete A1c indicator ($<6\%\textrm{--reference level}, 6-6.5\%, 6.5-7\%, 7-7.5\%, 7.5-8\%, 8-9\%,  \geq 9\%$), where the levels are chosen to coincide with those used by ADA to translate A1c into blood glucose level~\citep{diabetes18-target}. For simplicity, we do not add other factors.  For each imputation, we estimate the incidence from logistic regression models and account for the correlation between repeated measures of the same patient by implementing generalized estimating equations with an exchangeable working correlation structure. We apply MI combining rules~\citep{rubin:1987} to obtain the standard error estimate accounting for the uncertainty due to imputation. Figure~\ref{risk-pred} depicts the predicted incidence for individuals grouped by A1c levels from the four methods, and Table~\ref{risk-ci} presents both the point estimates and 95\% CIs of the predictions. Figure~\ref{risk-pred} shows that the relationship between A1c values and the risk of adverse outcomes is U-shaped.

Importantly, all analyses verify that A1c at or above 8\% is associated with a higher incidence of adverse events than levels of 6-7.5\% as expected from ADA recommendations to maintain level below 7\% or below 6.5\%, if possible. In fact, imputation does not notably change means for the two categories above 8\%, where less than 10\% of A1c values are imputed.  Results for levels $<6\%$, where approximately 20\% of values are imputed, are mixed, with BPMI as well as CCA showing a significantly higher incidence of acute events at this level of control than for A1c levels 6-7. 5\%. While MICE and MICE-2l also show higher incidence at A1c$<6\%$ than at the next two higher levels, these differences are not statistically significant.

\begin{table}[ht]
\centering
\caption{Predictive acute health event incidence risk and 95\% confidence intervals for individuals grouped by A1c levels. }
\label{risk-ci}
\begin{tabular}{rcccc}
  \hline
 & BPMI & MICE & MICE-2l & CCA \\ 
  \hline
$<6\%$ & 0.174 (0.164, 0.182) & 0.168 (0.16, 0.177) & 0.169 (0.161, 0.178) & 0.175 (0.164, 0.187) \\ 
${\bf 6 - 6.5\%}$ & 0.162 (0.156, 0.169) & 0.164 (0.157, 0.169) & 0.164 (0.157, 0.171) & 0.157 (0.149, 0.164) \\ 
${\bf 6.5 - 7\%}$ & 0.161 (0.154, 0.168) & 0.164 (0.157, 0.169) & 0.164 (0.157, 0.169) & 0.154 (0.147, 0.161) \\ 
$7 - 7.5\%$ & 0.164 (0.157, 0.172) & 0.165 (0.158, 0.172) & 0.165 (0.158, 0.172) & 0.157 (0.149, 0.165) \\ 
 ${\bf 7.5 - 8\%}$ & 0.165 (0.157, 0.175) & 0.167 (0.157, 0.175) & 0.165 (0.156, 0.174) & 0.155 (0.146, 0.165) \\ 
$8 - 9\%$ & 0.178 (0.169, 0.187) & 0.178 (0.169, 0.187) & 0.177 (0.167, 0.185) & 0.175 (0.164, 0.185) \\ 
$9\%+$ & 0.193 (0.179, 0.206) & 0.192 (0.179, 0.206) & 0.192 (0.179, 0.206) & 0.196 (0.182, 0.211) \\ 
   \hline
\end{tabular}
\end{table}

The incidence between A1c of 6\% and 8\% shows the largest differences between CCA and analyses with imputed data.  MICE and MICE-2l agree with BPMI in showing higher incidence than CCA in this range, but BPMI shows a slightly lower incidence than the other two methods, indicating that BPMI tends to place more low risk individuals in this category. The evidence from BPMI is similar to the finding in~\cite{a1c-death2019}. The fact that all methods show higher incidence than CCA in this range may be due to a complex interplay of reasons for missed A1c tests and their consequences, where missed A1c may be a marker for either good and stable glycemic control, or for lower utilization of preventive care. Future work will study the heterogeneous effects of patent complexity in the nonlinear relationship.

It seems that the probability of missingness depends on the outcome in this example, so that the missingness induced by A1c is MNAR in the acute event model. However, it is less clear if explicitly taking the outcome into account would change the relationship between imputed A1c and the outcome. Doing so would require specifying a correct model for the relationship between A1c and the outcome. Including the outcome as a covariate with main effects in MICE-2l, made no difference in predictions. We posit that imputing the A1c from a rich set of repeated measures for each individual minimizes the information the outcome can bring to the imputation. It should also be noted that baseline event incidence was included as a covariate in the imputation.

\subsubsection{Profiling heterogeneity and interpretation}

%\begin{table}[t]
%\centering
%\caption{Profile specific estimates for scale and coefficients of spline functions of time.}
%\label{L3-sigma-alpha}
%\footnotesize
%\begin{tabular}{lllc}
%  \hline
% &\#Patients &Scale (95\% CI) & Coefficients of spline functions (95\% CI) \\ 
%  \hline
% Profile 1 (good control)&3694& 0.29 (0.28, 0.29)&0\\
% \hline
% Profile 2 (fair control)  &2682&0.65 (0.64, 0.65)& \shortstack{0.85(0.8,0.9),  0.77(0.67,0.86), 0.76(0.69,0.82), \\0.85(0.79,0.91), 0.93(0.87,0.99), 0.9(0.83,0.97), \\ 0.91(0.81,1), 1.02(0.91,1.13), 0.97(0.86,1.07)}\\ 
% \hline
% Profile 3  (bad control)  &996 &1.42 (1.39, 1.45)  &\shortstack{ 2.35(2.21,2.49), 2.22(1.91,2.52), 2.23(2.04,2.43), \\2.39(2.23,2.56), 2.17(1.97,2.36), 2.23(2.03,2.44), \\2.38 (2.07,2.69), 2.42(2.02,2.83), 2.33(1.95,2.72)}\\ 
%   \hline
%\end{tabular}
%\end{table}

\begin{figure}
\begin{tabular}{cc}
\includegraphics[width=0.475\textwidth]{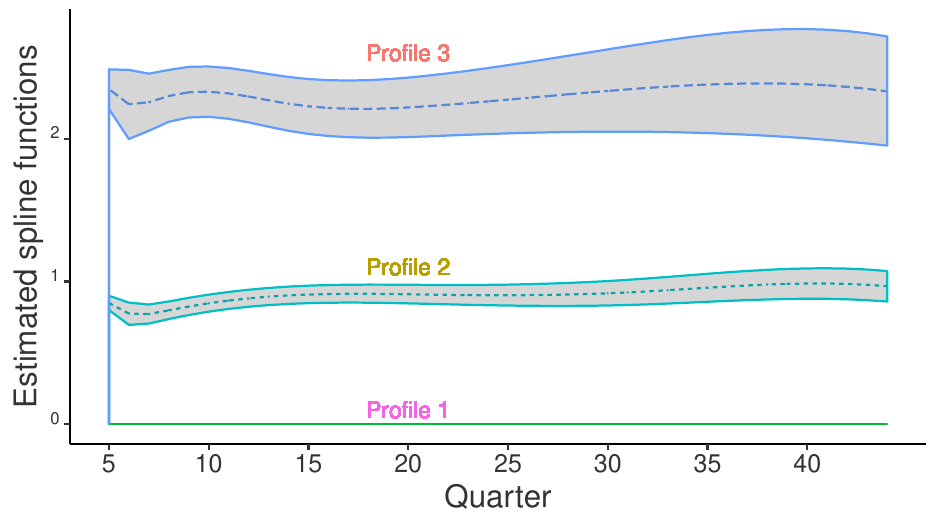}&\includegraphics[width=0.4755\textwidth]{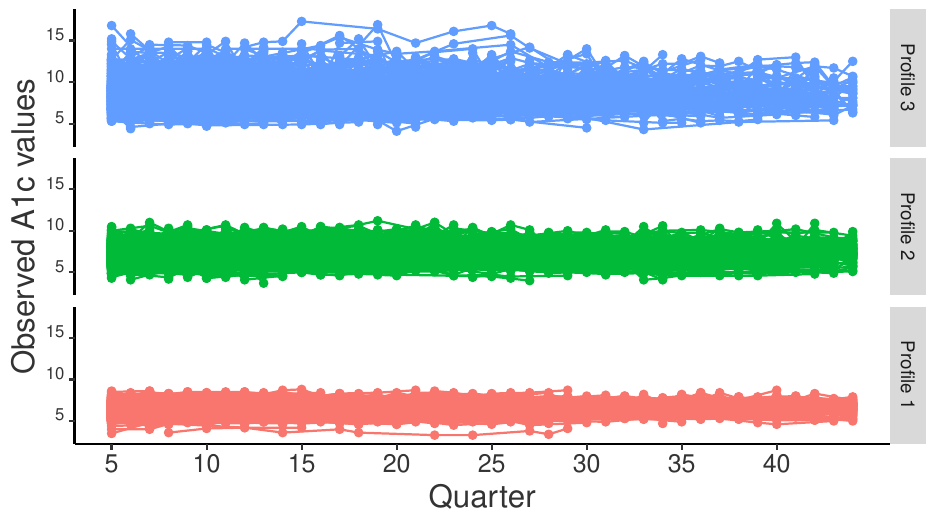}\\
\end{tabular}
\caption{Estimated spline functions with 95\% confidence intervals and observed A1c values of patients across profiles.}
\label{A1c_class}
\end{figure}

\begin{table}[t]
\centering
%\small
\caption{Descriptive statistics of baseline characteristics (first 4 quarters) and their coefficients for allocation across profiles.}
\label{baseline_class_con}
\begin{tabular}{lccc}
  \hline
 & Profile 1 & Profile 2 & Profile 3 \\ 
  \hline
  \multicolumn{4}{c}{Descriptive statistics: mean (sd)}\\
\#low prevalence& 0.22 (0.54) & 0.21 (0.54) & 0.21 (0.54) \\ 
\#ER visits & 0.11 (0.39) & 0.10 (0.29) & 0.17 (0.45) \\ 
\#Adverse events & 0.54 (0.85) & 0.57 (0.86) & 0.73 (0.97) \\ 
  HCC & 1.26 (0.84) & 1.57 (1.11) & 1.57 (1.01) \\ 
  Age & 71.52 (9.98) & 69.76 (10.28) & 64.33 (12.73) \\ 
  A1c & 6.38 (0.63) & 7.23 (0.93) & 8.5 (1.66) \\ 
   \hline
     \multicolumn{4}{c}{Coefficients: log-odds (95\% CI)}\\
   Intercept &0& -6.08 (-7.51, -4.94) & -9.15 (-10.29, -8.01) \\ 
\#low prevalence&0  &  -0.34 (-0.52, -0.16) & -0.44 (-0.68, -0.21) \\ 
\#ER visits &0& -0.46 (-0.81, -0.13) & -0.41 (-0.79, -0.05) \\ 
\#Adverse events& 0&0.12 (-0.01, 0.25) & 0.15 (-0.02, 0.32) \\ 
  HCC &0& 0.58 (0.36, 0.73) & 0.59 (0.36, 0.76) \\ 
  Age &0 &-0.06 (-0.07, -0.04) & -0.11 (-0.12, -0.09) \\ 
  Baseline A1c &0 &1.35 (1.25, 1.46) & 2.06 (1.92, 2.19) \\ 
  \hline
\end{tabular}
\end{table}

As one important byproduct of BPMI, the latent profiling provides an interpretable summary of patient heterogeneity in A1c trajectories. As an illustration, we randomly draw the latent class indicators based on the posterior allocation probabilities and use the last draw for the profiling assignment. The three profiles have different trajectories over time and different scales. Figure~\ref{A1c_class} shows the estimated spline functions and 95\% CIs, as well as the observed A1c values across profiles. We present the two plots in Figure~\ref{A1c_class} to illustrate the profiling structure, but they are not directly comparable. The spline functions ($\bfD_i^*\bfbeta^*_{C_i}$) are part of the mean structure, so the two plots have different y-axis ranges. Meanwhile, we have included additional time-varying variables, and they both have significantly positive relationships with A1c values, where the coefficient estimate of age is 0.007 with 95\% CI (0.006, 0.008), and the coefficient estimate of the count of physician visits is 0.003 with 95\% CI (0.001, 0.006).

The profile allocation assigns 3694 patients to the largest class with the smallest variability of residuals, 0.29. The smallest class has 996 patients and the largest variability of A1c values (1.42), with large coefficients of the spline functions. The remaining class has 2692 patients, and the estimated variability of A1c values is 0.65. The patients within Profile 1 are under good control with low A1c levels and stability, of whom all the A1c values are below 9\%. Patients within Profile 2 are under fair control with modest A1c values, while those with Profile 3 are under bad control with high A1c levels and variability.

\begin{figure}
\begin{tabular}{cc}
\includegraphics[width=0.5\textwidth,height=3.5in]{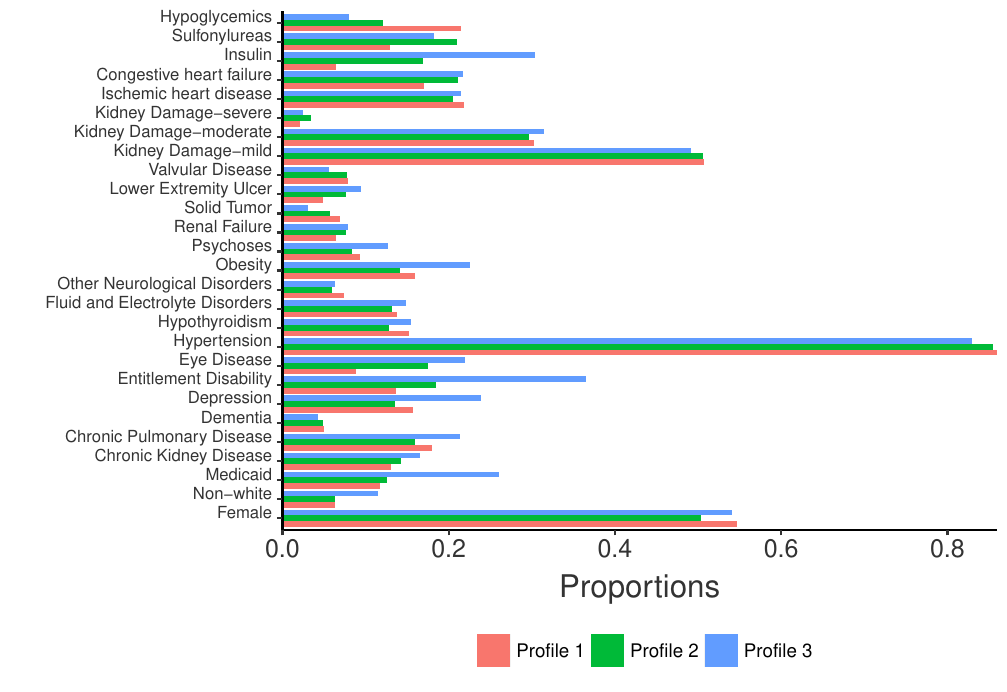}&
\includegraphics[width=0.5\textwidth,height=3.5in]{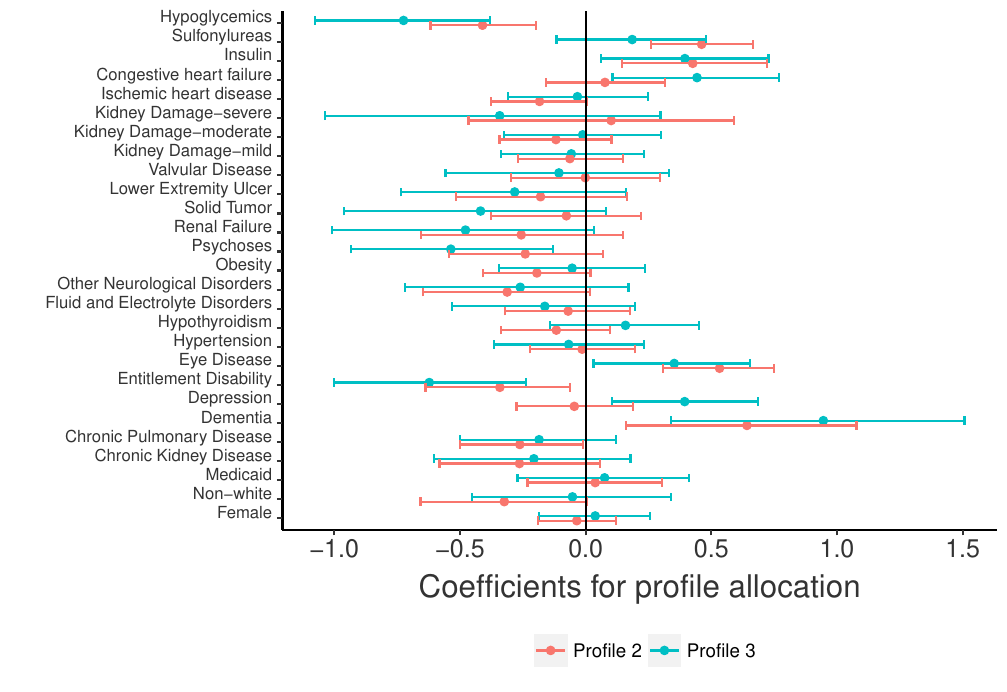}\\
\end{tabular}
\caption{Frequency distribution of categorical baseline characteristics and their coefficients for profile allocation.}
\label{baseline_class_cat}
\end{figure}

The profiling is primarily dependent on the A1c trajectories, and the profile allocation probabilities are predicted by the baseline covariates. Figure~\ref{baseline_class_cat} and Table~\ref{baseline_class_con} provide the descriptive summary of the baseline characteristics and the corresponding coefficients for profile allocation. The patients in poor control in Profile 3 tend to be non-white, have Medicaid coverage and more complications (such as Depression, Entitlement Disability, Eye disease, Psychoses and Obesity), and take Insulin. These patients also have more ER visits, more adverse events, higher HCC and baseline A1c values than those in good/fair control.

The coefficients for allocation across profiles give the logarithm of the relative odds to be assigned to other profiles compared to Profile 1. With higher baseline HCC and A1c values, the patients tend to be out of good control. Non-white patients with Chronic Pulmonary Disease, Entitlement Disability, other neurological disorders, Obesity, Psychoses and taking the hypoglycemic drug, have significantly higher odds of being out of good control. The patients with Dementia, Eye disease, congestive heart failure and those who take Insulin or Sulfonylureas tend to have good A1c control.

\section{Discussion}
\label{discussion}

Electronic health records play an increasingly important role in evidence-driven research on effective approaches to clinical therapy. However, as data are collected in an unstructured manner when patient contacts occur, EHR suffers from missing data, where for example, outcomes and predictors are not always available in the same time interval. MI is an attractive approach for filling in data needed for various analyses, allowing assessment of imputation uncertainty. However, few methods exist for filling in intermittently missing data in long time sequences. In addition, the missing mechanism is often complex raising the possibility of informative missingness. We develop a method for multiple imputation based on Bayesian latent profiling, which allows for missingness being intermittent and potentially MNAR. The method allows for a flexible mixture distribution by allowing location and scale parameters to vary across latent profiles, covariates to predict profile membership and a time trend to be modeled by splines. We compare BPMI with popular MI MAR based approaches MICE and MICE-2l by simulations in a range of settings, and find BPMI to be the most unbiased and efficient. We apply our method to missing A1c in EHRs, and find BPMI to perform well with 3 latent classes, in terms of BIC, LPML and posterior predictive checks. Hence, the theoretical properties for our approach seem to be satisfactory.

As we predict acute adverse outcomes from the EHR with and without imputation, BPMI as well as MICE and MICE-2l predicts more adverse events than does CCA in the 7-8\% A1c range. Hence, CCA analysis of the EHR data cannot be trusted. While a MAR structure fits the A1c data, it appears that the missingness induced in the outcome by missing A1c is not MAR. However, all outcomes are restored when A1c is imputed. Our future work will explore including the acute outcomes in the imputation as generally recommended. However, it is unclear how much this would change the imputed values given the availability of rich A1c trajectories and many covariates. It is also challenging to specify the model to properly account for the non-linear relationship and patient heterogeneity. Our approach avoids the need to model the dependence of the outcome on A1c, and improves the availability of the A1c values for investigating other outcomes. Importantly, our method fits a very flexible mixture model to capture the data distribution.

Several questions regarding the application of our method arise. We considered the case when only A1c values are missing and include numerous covariates. In practice, EHRs will suffer from incomplete information beyond A1c, such as the baseline variables in Table~\ref{predictor}. With multivariate incomplete variables, joint imputation or sequentially conditional imputation is necessary. It is also uncertain how many latent classes are needed. BPMI makes the crucial assumption of conditional independence, given the latent classes. Likely, a greater number of classes makes this assumption more plausible, but leads to difficulties including covariates, especially those with low prevalence. Our diagnostics does not find evidence against the assumed model. Future work will investigate the specific role of different aspects of patient complexity on the relationship of A1c to acute outcomes. This will address substantive questions as well as further illuminate the plausibility of the imputations.

Several extensions of BPMI may be of interest. Besides the incorporation of time-varying health outcomes into modeling, other time-varying variables including BP, BMI, and LDL can provide additional information to impute the A1c levels. To handle their high missingness percentages, we can simultaneously model the trajectories and impute missing values via joint mixture membership profiling~\citep{MMbook19}. The number of latent profiles can be treated as random in future work. With more information available from EHR, systematic variable selection in an integrative inference framework needs further investigation. For longitudinal data with intermittent missingness and nonlinear trajectories, it would be useful to connect our joint modeling approach with existing approaches for monotone missingness or weighting adjustments~\citep{hogandaniels,msm:robins00}.

%%%%%%%%%%%%%%%%%%%%%%%%%%
\section*{Acknowledgements}
This work is funded by the National Institutes of Health (R21DK110688 and R01DK108073), the Health Innovation Program, the UW School of Medicine and Public Health from The Wisconsin Partnership Program, the Community-Academic Partnerships core of the University of Wisconsin Institute for Clinical and Translational Research through the National Center for Advancing Translational Sciences, grant UL1TR000427, and the Agency for Healthcare Research and Quality (R01 HS018368 and R21 HS017646). The content is solely the responsibility of the authors and does not necessarily represent the official views of the NIH.

The authors thank Roderick Little for his constructive comments and help in the paper editing.
%%%%--------------------------------------------------------%%%

%\newpage

\bibliography{/Users/yajuan/box/bibs/ys-2020} 
\bibliographystyle{chicago} 
%%%%%--------------------------------------------------------%%%
%\newpage
\appendix
\section{Spline functions}
\label{spline}

\begin{figure}[h!]
\begin{tabular}{c}
\includegraphics[width=0.9\textwidth]{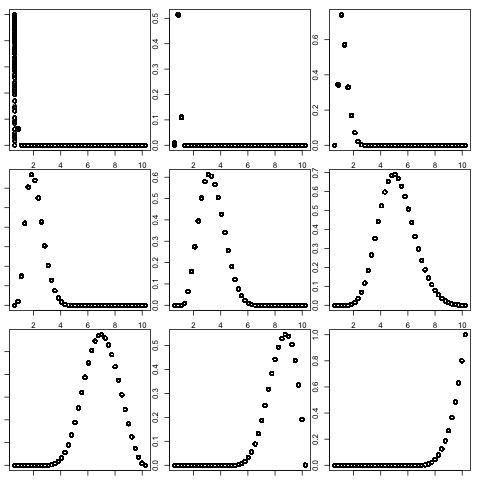}
\end{tabular}
\caption{Spline functions of time used as covariates in $\bfD^*_{ij}$ with profile-specific coefficients. The x-axis represents time in years.}
\label{bspline}
\end{figure}

\section{Posterior computation}
\label{posterior}
Supplemental materials for the posterior computation under marginal and joint profiling.

\subsection{Marginal profiling}
\label{ps-ig}
The prior distributions are specified as: $\beta\sim N(0,\mathbf{\Sigma}_{\beta})$, $\alpha_l \sim N(0,\mathbf{\Sigma}_{\alpha})$, $\sigma_l^{2}\sim \textrm{Inverse-Gamma: IG}(a,b)$, $\mathbf{\Sigma}_{r} \sim \textrm{Inverse-Wishart: IW}(\nu_b,\mathbf{\Sigma}_b)$ and $\eta_l\sim N(b_l, B_l)$, where we set the hyper-parameters equal to 1 and assume independence in the covariance matrix specification. The posterior updating follows the steps:
\begin{enumerate}
    \item Update $C_i$: for $i=1,\dots,n$, draw $C_i$ from a multinomial distribution with probability
    \begin{align*}
    \textrm{Pr}(C_i=l)=\frac{\pi_l(\bfX_{i0},\eta_l)\prod_jf(Y_{ij}|\beta,\alpha_l,b_i,\sigma_l^2,D_i,D_i^*,D_i^{**})}{\sum_{l} \pi_l(\bfX_{i0},\eta_l)\prod_jf(Y_{ij}|\beta,\alpha_l,b_i,\sigma_l^2,D_i,D_i^*,D_i^{**})}
    \end{align*}
    \item Update $\alpha_l$: for $l=2,\dots,L$, where $\alpha_1=0$ for identification purpose, 
    $\pi(\alpha_l|-)%&\propto \prod_{i=1}^n[\prod_{j=1}^{T_i}f(Y_{ij}|\beta,\alpha_l,b_i,\sigma_l^2,D_i,D_i^*,D_i^{**})]^{I_{(C_i=l)}}\\
    %&=\prod_{i: C_i=l}\prod_{j=1}^{T_i}f(Y_{ij}|\beta,\alpha_l,b_i,\sigma_l^2,D_i,D_i^*,D_i^{**})\\
    =N(\mu_{\alpha},V_{\alpha}),
    $
    where $V_{\alpha_l}=(\Sigma^{-1}_{\alpha}+\sigma_l^{-2}D_l^{*'}D_l^*)^{-1},$ $\mu_{\alpha}=V_{\alpha_l}\sigma_l^{-2}D^{*'}_l(X_{l}-D_l\beta-D_l^{**}b_l).$
    Here $D_l, D^*_l, D^{**}_l, X_l,b_l$ are subsets of the design matrix and random effects for all $i,j$ such that $C_i=l$
    
    \item Update $\eta_l$: for $l=2,\dots,L$, let $c_{il}=I(C_i=l)$,
%    \begin{align*}
%        \pi(\eta_l|-)&\propto \prod_{i=1}^n[\textrm{Pr}(C_i=l|\pi_{l}(\bfX_{i0},\eta_l))]^{I_{(C_i=l)}}\pi(\eta_l)\\
%                   &\propto\prod_{i=1}^n\frac{\exp(\bfX_{i0}^T\eta_lc_{il})}{\sum_{k=1}^L\exp(\bfX_{i0}^T\eta_k)}\pi(\eta_l)\\
%                   &\propto\prod_{i=1}^n\frac{\exp(r_{il})^{c_{il}}}{1+\exp(r_{il})}\pi(\eta_l),
%    \end{align*}
%where $r_{il}=\bfX_{i0}^T\eta_l-\log[\sum_{k\neq l}\exp(\bfX_{i0}^T\eta_k)]$. The P\'{o}lya-Gamma  random variables $w_{il}|\bfX_{i0},\eta\sim PG(1,r_{il})$ are independent of all the rest given $\eta, \bfX_{i0}$. We have
%\[
%\frac{\exp(r_{il})^{c_{il}}}{1+\exp(r_{il})}\propto \exp(r_{il}(c_{il}-1/2))\int_{0}^{\infty}\exp\{-w_{il}r^2_{il}/2\}PG(w_{il}\mid1,0)dw_{il},
%\]
%where $PG(w_{il}\mid 1,0)$ is the P\'{o}lya-Gamma prior distribution with parameters (1,0) for $w_{il}$. Hence the conditional posterior distribution of $\eta_l$ given $w$ is
%\begin{align*}
$\pi(\eta_l)
%&\propto \prod_{i=1}^n\exp(r_{il}(c_{il}-1/2))\exp\{-w_{il}r^2_{il}/2\}N(b_l,B_l)\\
\sim N(m^*_{l},S^*_{l})$,
%\end{align*}
where 
$S^*_{l}=(V^T\Omega_{l}V+B_l^{-1})^{-1}$, $V$ is the design matrix with each row $\bfX_{i0}^T$, $\Omega_{l}=diag(w_{1l},\dots,w_{nl})$, and
$
m^*_l=S^*_l(B_l^{-1}b_l+V^Tm_l),
$
where $m_l$ is a vector in $R^n$, and the $i$th component is 
%$\begin{align*}
$
m_l^{(i)}=c_{il}-1/2 + w_{il}\{\log[\sum_{k\neq l}\exp(\bfX_{i0}^T\eta_k)]\}.
$
%\end{align*}
%Hence updating $\eta_l$ involves two steps: $(w_{il}|-)\sim PG(1,r_{il})$ and $(\eta_l|-)\sim N(m^*_{l},S^*_{l})$.
% BayesLogit rpg
    \item Update $\beta\sim N(\mu_{\beta}, V_{\beta})$,where
    %\begin{align*}
    %\pi(\beta|-)&\propto \prod_{i=1}^n\prod_{j=1}^{T_i}f(Y_{ij}|\beta,\alpha_{C_i},b_i,\sigma_{C_i}^2,D_i,D_i^*,D_i^{**})\pi(\beta).
    %\end{align*}
    %Integrate the random effect $b_i$ out,
    %\[Y_i\sim N(D_i\beta+D^*_i\alpha_{C_i}, D_i^{**}\mathbf{\Sigma}_rD_i^{**'}+\sigma_{C_i}^2I_{T_i}),\]
    $\bfX_{i0}^*=D_i^{**}\mathbf{\Sigma}_rD_i^{**'}+\sigma_{C_i}^2I_{T_i}$,
    $V_{\beta}=(\mathbf{\Sigma}_{\beta}^{-1}+\sum_iD_i'{\bfX_{i0}^*}^{-1}D_i)^{-1}$,
    and
    $\mu_{\beta}=V_{\beta}\sum_iD_i'{\bfX_{i0}^*}^{-1}(Y_i-D^*_i\alpha_{C_i})$. 
    
    \item Update $\sigma_l^2$, for $l=1,\dots, L$ from
    $
    IG(\sum_{i:C_i=l}T_i/2+a, (Y_l-D_l\beta-D_l^*\alpha-D_l^{**}b_l)'(Y_l-D_l\beta-D_l^*\alpha-D_l^{**}b_l)/2+b).
    $
    %where $D_l, D^*_l, D^{**}_l, Y_l,b_l$ are subsets of the data and random effects for all $i,j$ such that $C_i=l$
    \item Update $\mathbf{\Sigma}_r$ from
    $IW(n+\nu_b,b'b+\mathbf{\Sigma}_b)$.
    \item Update $b_i$
    from
    $N(\mu_b,V_b)$,
    here $V_b=(\mathbf{\Sigma}_r^{-1} + \sigma_{C_i}^{-2}D_i^{**'}D_i^{**})^{-1},$ and $\mu_b=V_b\sigma_{C_i}^{-2}D_i^{**'}(Y_i-D_i\beta-D_i^{*}\alpha_{C_i})$
    \end{enumerate}
After convergence, we impute missing data from the assumed model for $Y$ since this step does not need to be included into the iterations.

\subsection{Joint profiling}
\label{ps-nonig}

The prior distributions are specified as $\beta\sim N(0,\mathbf{\Sigma}_{\beta})$, $\alpha_l \sim N(0,\mathbf{\Sigma}_{\alpha})$, $\sigma_l^{2}\sim \textrm{IG}(a,b)$, $\mathbf{\Sigma}_{r} \sim \textrm{IW}(\nu_b,\mathbf{\Sigma}_b)$, $\eta_l\sim N(b_l, B_l)$, $\nu\sim N(0,\mathbf{\Sigma}_{\nu})$, $\gamma_l \sim N(0, \mathbf{\Sigma}_{\gamma})$ and $E \sim IW(\nu_e, \mathbf{\Sigma}_e)$,
where we set the hyper-parameters equal to 1 and assume independence in the covariance matrix specification. The posterior computation is the following.
\begin{enumerate}
    \item Update $C_i$: for $i=1,\dots,n$, draw $C_i$ from a multinomial distribution with probability
    \begin{align*}
    \textrm{Pr}(C_i=l)=\frac{\pi_l(\bfX_{i0},\eta_l)\prod_jp_{ij|C_i=l}^{R_{ij}}(1-p_{ij|C_i=l})^{1-R_{ij}}f(Y_{ij}|\beta,\alpha_l,b_i,\sigma_l^2,D_i,D_i^*,D_i^{**})}{\sum_{l} \pi_l(\bfX_{i0},\eta_l)\prod_jp_{ij|C_i=l}^{R_{ij}}(1-p_{ij|C_i=l})^{1-R_{ij}}f(Y_{ij}|\beta,\alpha_l,b_i,\sigma_l^2,D_i,D_i^*,D_i^{**})}.
    \end{align*}

    \item Update $\alpha_l$: for $l=2,\dots,L$, where $\alpha_1=0$, 
    %\begin{align*}
    $\pi(\alpha_l|-)%&\propto \prod_{i=1}^n[\prod_{j=1}^{T_i}f(Y_{ij}|\beta,\alpha_l,b_i,\sigma_l^2,D_i,D_i^*,D_i^{**})]^{I_{(C_i=l)}}\\
    %&=\prod_{i: C_i=l}\prod_{j=1}^{T_i}f(Y_{ij}|\beta,\alpha_l,b_i,\sigma_l^2,D_i,D_i^*,D_i^{**})\\
    =N(\mu_{\alpha},V_{\alpha})$,
    where $V_{\alpha_l}=(\Sigma^{-1}_{\alpha}+\sigma_l^{-2}D_l^{*'}D_l^*)^{-1},$ $\mu_{\alpha}=V_{\alpha_l}\sigma_l^{-2}D^{*'}_l(Y_{l}-D_l\beta-D_l^{**}b_l).$
    
    \item Update $\eta_l$: for $l=2,\dots,L$, % update as a vector $\eta_1,\dots,\eta_L$. Let $c_{il}=I(C_i=l)$,
%    \begin{align*}
%        \pi(\eta_l|-)&\propto \prod_{i=1}^n[\textrm{Pr}(C_i=l|\pi_{l}(\bfX_{i0},\eta_l))]^{I_{(C_i=l)}}\pi(\eta_l)\\
%                   &\propto\prod_{i=1}^n\frac{\exp(\bfX_{i0}^T\eta_lc_{il})}{\sum_{k=1}^L\exp(\bfX_{i0}^T\eta_k)}\pi(\eta_l)\\
%                   &\propto\prod_{i=1}^n\frac{\exp(r_{il})^{c_{il}}}{1+\exp(r_{il})}\pi(\eta_l),
%    \end{align*}
%where $r_{il}=\bfX_{i0}^T\eta_l-\log[\sum_{k\neq l}\exp(\bfX_{i0}^T\eta_k)]$ and random variables $w_{il}|\bfX_{i0},\eta\sim PG(1,r_{il})$, independent of all the rest given $\eta, \bfX_{i0}$. We have
%\[
%\frac{\exp(r_{il})^{c_{il}}}{1+\exp(r_{il})}\propto \exp(r_{il}(c_{il}-1/2))\int_{0}^{\infty}\exp\{-w_{il}r^2_{il}/2\}PG(w_{il}\mid1,0)dw_{il},
%\]
%where $PG(w_{il}\mid 1,0)$ is the P\'{o}lya-Gamma prior distribution with parameters (1,0) for $w_{il}$. Hence the conditional posterior distribution of $\eta_l$ given $w$ is
$%\begin{align*}
\pi(\eta_l)%&\propto \prod_{i=1}^n\exp(r_{il}(c_{il}-1/2))\exp\{-w_{il}r^2_{il}/2\}N(b_l,B_l)\\
\sim N(m^*_{l},S^*_{l})$,
where 
$S^*_{l}=(V^T\Omega_{l}V+B_l^{-1})^{-1}$, $V$ is the design matrix with each row $\bfX_{i0}^T$, $\Omega_{l}=diag(w_{1l},\dots,w_{nl})$, and
$
m^*_l=S^*_l(B_l^{-1}b_l+V^Tm_l)$,

where $m_l$ is a vector in $R^n$, and the $i$th component is 
\begin{align*}
m_l^{(i)}=c_{il}-1/2 + w_{il}\{\log[\sum_{k\neq l}\exp(\bfX_{i0}^T\eta_k)]\}
\end{align*}
%Hence updating $\eta_l$ involves two steps:
%\begin{enumerate}
%    \item $(w_{il}|-)\sim PG(1,r_{il})$
%    \item $(\eta_l|-)\sim N(m^*_{l},S^*_{l})$.
%\end{enumerate}
% BayesLogit rpg
    \item Update $\beta$: $\pi(\beta|-)=N(\mu_{\beta}, V_{\beta})$,
%    \begin{align*}
%    \pi(\beta|-)&\propto \prod_{i=1}^n\prod_{j=1}^{T_i}f(Y_{ij}|\beta,\alpha_{C_i},b_i,\sigma_{C_i}^2,D_i,D_i^*,D_i^{**})\pi(\beta)\\
%    &=N(\mu_{\beta}, V_{\beta}),
%    \end{align*}
%Integrate the random effect $b_i$ out,
%    $Y_i\sim N(D_i\beta+D^*_i\alpha_{C_i}, D_i^{**}\mathbf{\Sigma}_rD_i^{**'}+\sigma_{C_i}^2I_{T_i}),\]
%    and let 
$\bfX_{i0}^*=D_i^{**}\mathbf{\Sigma}_rD_i^{**'}+\sigma_{C_i}^2I_{T_i}$,
    $V_{\beta}=(\mathbf{\Sigma}_{\beta}^{-1}+\sum_iD_i'{\bfX_{i0}^*}^{-1}D_i)^{-1}$,
    and
    $\mu_{\beta}=V_{\beta}\sum_iD_i'{\bfX_{i0}^*}^{-1}(Y_i-D^*_i\alpha_{C_i})$. 
%    Use Cholesky decomposition of $\bfX_{i0}^*=L_iL'_i$, where $L_i$ is a lower triangular matrix, then
%    $
%    \sum_iD_i'{\bfX_{i0}^*}^{-1}D_i=\sum_iD_i'(L^{-1}_i)'L^{-1}_iD_i=(L^{-1}D)'L^{-1}D,
%    \]
%    where $(L^{-1}D)'=((L_1^{-1}D_1)',\dots,(L_n^{-1}D_n)')$. Similarly,
%    $
%    \sum_iD_i'{\bfX_{i0}^*}^{-1}(Y_i-D^*_i\alpha_{C_i})=\sum_iD_i'(L^{-1}_i)'L^{-1}_i(Y_i-D^*_i\alpha_{C_i})=(L^{-1}D)'L^{-1}Y^*,
%    \]
%    where $(L^{-1}Y^*)'=((L_1^{-1}(Y_1-D^*_1\alpha_{C_1}))',\dots,(L_n^{-1}(Y_n-D^*_n\alpha_{C_n}))')$.
    \item Update $\sigma_l^2$, for $l=1,\dots, L$, from 
$
    IG(\sum_{i: C_i=l}T_i/2+1, (Y_l-D_l\beta-D_l^*\alpha-D_l^{**}b)'(Y_l-D_l\beta-D_l^*\alpha-D_l^{**}b)/2+1).
$
%    where $D_l, D^*_l, D^{**}_l, Y_l,b_l$ are subsets of the data and random effects for all $i,j$ such that $C_i=l$.
    \item Update $\mathbf{\Sigma}_r$ from $IW(n+\nu_b,b'b+\mathbf{\Sigma}_b)$.
    \item Update $b_i$ from
    $\pi(b_i|-)%&\propto\prod_{j=1}^{T_i}f(Y_{ij}|\beta,\alpha_{C_i},b_i,\sigma_{C_i}^2,D_i,D_i^*,D_i^{**})f(b_i|\mathbf{\Sigma}_{r\times r})\\
    =N(\mu_b,V_b)$,
    here $V_b=(\mathbf{\Sigma}_r^{-1} + \sigma_{C_i}^{-2}D_i^{**'}D_i^{**})^{-1},$ and $\mu_b=V_b\sigma_{C_i}^{-2}D_i^{**'}(Y_i-D_i\beta-D_i^{*}\alpha_{C_i}).$
\item Update $\nu$:
$[v|-] \sim N(m_{v}^{*}, S_{v}^{*})$,
where $m_{v}^{*}= S_{v}^{*}(B^{\top}k_{v})$, 
$%\begin{align*}
B=
\left[ \begin{array}{c} B_{11}^{\top}, ... ,B_{1T_{1}}^{\top},B_{21}^{\top},...,B_{n T_{n}}^{\top} \end{array} \right] ^{\top}
$%\end{align*}
and $k_{v}$ is a vector with length $\sum_{i=1}^{n}T_{i}$:
$$
k_{v}=
\left[ \begin{array}{c} k_{11}^{*}-w_{11}^{*}((B_{11}^{*})^{\top}\gamma_{C_1}+(B_{11}^{**})^{\top}e_{1})),...,k_{nT_{n}}^{*}-w_{nT_{n}}^{*}((B_{nT_n}^{*})^{\top}\gamma_{C_n}+(B_{nT_n}^{**})^{\top}e_{n})) \end{array} \right]^{\top}
$$
and 
$
S_{v}^{*}=(\mathbf{\Sigma}_{v}^{-1}+B^{\top}\Omega_{v}B)^{-1},
$
where $\Omega_{v}=diag( w_{11}^{*},..., w_{1T_{1}}^{*}, w_{21}^{*},...,w_{2T_{2}}^{*},...,w_{n1}^{*},..., w_{nT_{n}}^{*})$. 
\item Update $\gamma_l$, for $l=2,\dots, k$, from
$[\gamma_l|-] \sim N(m_{\gamma_l}^{*}, S_{\gamma_l}^{*})$,
where $m_{\gamma_l}^{*}= S_{\gamma_l}^{*}((B^*)^{\top}k_{\gamma_l})$, 
$
B^*_l=
\left[ \begin{array}{c} (B^*_{11})^{\top}I(C_1=l), ..., (B^*_{1T_{1}})^{\top}I(C_1=l), (B^*_{21})^{\top}I(C_2=l),...,B^*_{n T_{n}})^{\top}I(C_n=l) \end{array} \right]^{\top}
$
and $k_{\gamma_l}$ is a vector with length $\sum_{i:C_i=l}T_{i}$:
\begin{align*}
k_{\gamma_l}=
\left[ \begin{array}{c} \{k_{11}^{*}-w_{11}^{*}(B_{11}^{\top}\nu+(B_{11}^{**})^{\top}e_{1})\}I(C_1=l),...,\{k_{nT_{n}}^{*}-w_{nT_{n}}^{*}(B_{nT_n}^{\top}\nu+(B_{nT_n}^{**})^{\top}e_{n})\}I(C_n=l)  \end{array} \right]^{\top}
\end{align*}
and 
$
S_{\gamma_l}^{*}=(\mathbf{\Sigma}_{\gamma_l}^{-1}+(B^*)^{\top}\Omega_{\gamma_l}B^*)^{-1},
$
where $\Omega_{\gamma_l}=diag( w_{11}^{*}I(C_1=l),..., w_{1T_{1}}^{*}I(C_1=l), w_{21}^{*}I(C_2=l),...,w_{2T_{2}}^{*}I(C_2=l),...,w_{n1}^{*}I(C_n=l),..., w_{nT_{n}}^{*}I(C_n=l))$. 
\item Update $e_i$ from
$[e_{i}|-] \sim N(m_{e_{i}}^{*}, S_{e_{i}}^{*})$,
where $m_{e_{i}}^{*}=S_{e_{i}}^{*}((B^{**}_{i})^{\top}k_{e_{i}})$, 
$
B^{**}_{i}=
\left[ \begin{array}{c} (B_{i1}^{*})^{\top}, ... ,(B_{iT_{i}}^{*})^{\top} \end{array} \right]^{\top}
$
and $k_{e_{i}}$ is a vector of $R^{T_{i}}$ as the following:
\begin{align*}
k_{e_{i}}=
\left[ \begin{array}{c} k_{i1}^{*}-w_{i1}^{*}(B_{i1}^{\top}\nu+(B_{i1}^{*})^{\top}\gamma_{C_1}),...,k_{iT_{i}}^{*}-w_{iT_{i}}^{*}(B_{iT_i}^{\top}\nu+(B_{iT_i}^{*})^{\top}\gamma_{C_i}) \end{array} \right]^{\top}
\end{align*}
and 
$
S_{e_{i}}^{*}=(E^{-1}+((B^{**}_{i})^{\top}\Omega_{e_{i}}B^{**}_{i})^{-1},
$
where $\Omega_{e_{i}}=diag( w_{i1}^{*},..., w_{iT_{i}}^{*})$. 

\item Update $E$ from 
$
IW(n+\nu_{e}, \sum_{i=1}^{n}e_{i}e_{i}^{\intercal}+\mathbf{\Sigma}_{e}).
$
\item Impute missing data: draw $Y_{ij}^*$ from $f(Y_{ij}|\beta,\alpha_{C_i},b_i,\sigma_{C_i}^2,D_i,D_{ij}^*,D_i^{**})$ .
\end{enumerate}

\end{document}